\documentclass[3p,11pt]{elsarticle}

\pdfoutput=1

\usepackage{epsfig,graphicx,amssymb,amsmath, amsthm,subeqnarray,rotating,color,float}

\graphicspath{{figs/}{figs_lo/}}

\def\varepsilon{\epsilon}

\journal{Physica D: Nonlinear Phenomena}

\begin{document}

\begin{frontmatter}

\title{Microorganism Billiards}
\author{Saverio E. Spagnolie, Colin Wahl, Joseph Lukasik and Jean-Luc Thiffeault}
\address{Department of Mathematics, University of Wisconsin -- Madison\\
  480 Lincoln Dr., Madison, WI 53706}

\begin{abstract}
Recent experiments and numerical simulations have shown that certain types of microorganisms ``reflect'' off of a flat surface at a critical angle of departure, independent of the angle of incidence. The nature of the reflection may be active (cell and flagellar contact with the surface) or passive (hydrodynamic) interactions. We explore the billiard-like motion of a body with this empirical reflection law inside a regular polygon and show that the dynamics can settle on a stable periodic orbit or can be chaotic, depending on the swimmer's departure angle and the domain geometry. The dynamics are often found to be robust to the introduction of weak random fluctuations. The Lyapunov exponent of swimmer trajectories can be positive or negative, can have extremal values, and can have discontinuities depending on the degree of the polygon. A passive sorting device is proposed that traps swimmers of different departure angles into separate bins. We also study the external problem of a microorganism swimming in a patterned environment of square obstacles, where the departure angle dictates the possibility of trapping or diffusive trajectories.
\end{abstract}

\begin{keyword}
Billiards \sep Non-specular reflections \sep Microorganism locomotion \sep Sorting
\end{keyword}

\end{frontmatter}

\section{Introduction}
Microorganisms often follow unexpected trajectories when swimming near surfaces. Organisms such as spermatozoa and {\it E.~coli} are attracted to surfaces and may accumulate there due to a combination of hydrodynamic and steric effects \cite{Rothschild63,fm95,btbl08,sgbk09,sb09,ddcgg11,sl12}. This has biological and even medical implications, as attraction and accumulation may lead to the development of biofilms \cite{vllnz90,otkk00,kd10}, and infection of medically-implanted surfaces \cite{hdf92}. The orientations of swimming bodies in such an environment depend on the geometry of the swimmers, their mechanism of propulsion, and their interactions \cite{gnbnm05,sgs10,giy10,znm09,houg09,co10,lp10,Crowdy11,sl12,updt15}. In addition to being attracted to walls, \textit{E.~coli} cells tend to swim in large circular trajectories due to hydrodynamic interactions with the surface \cite{ldlws06}; notably, the orientation of the circular trajectories are reversed for swimming near a fluid-air interface \cite{lpcp10,dldaai11}, which can be rationalized by considering the two different image flows near no-slip and shear-free surfaces \cite{dldaai11,sl12}.

The unusual behaviors of microorganisms near surfaces has motivated the development of many intriguing engineering applications. Wedge-shaped obstacles have been designed to passively sort \textit{E.~coli} cells into areas of differing concentrations \cite{gkca07}, and \textit{E.~coli} cells have been used to drive micro-devices powered by asymmetric gears  \cite{adr09,dladariscmdadf10,saga10,lz13}. Sorting and rectification devices that exploit the interactions of microorganisms and asymmetric surfaces (including funnels and gears) have also been designed and studied \cite{gkca07,wrnr08,tc09,dladariscmdadf10,bjmvdvscm13}. In some cases, steric collisions or near-field lubrication forces may dominate long-range hydrodynamic effects \cite{ddcgg11,wwdkg13,lwg14}.

Swimming trajectories are naturally more intricate in these complex environments. For instance, Takagi~{\it et al.}~showed that microswimmers in a field of passive colloidal beads display a billiard-like motion between colloids, intermittent periods of entrapped, orbiting states near single colloids, and randomized escape behavior \cite{tpbsz14}. Brown~{\it et al.}~extended this work to swimming through a ``colloidal crystal,'' where a synthetic swimmer hops from colloid to colloid with a trapping time that depends on fuel concentration, while {\it E.~coli} trajectories are rectified into long, straight runs \cite{bvdvslp15}. Coupling topography to propulsive mechanism has also been exploited to guide active Janus particles \cite{skupts16}. A far-field hydrodynamic model predicts attraction, entrapment, and scattering of particles near spherical colloids \cite{smbl15}, though the nature of entrapment is generally specific to the microorganism or active particle \cite{smbl15,szs15,cltkp15,sndg15,weg15,mskm16,lbsm16}. {\it E.~coli}, for instance, can be trapped by convex walls in part because the cell body pitches down toward the surface in equilibrium \cite{sl12,sndg15}; moreover, the tumbling behavior of {\it E.~coli} is suppressed near surfaces due to increased hydrodynamic resistance \cite{trb00,mbss14}. For a review of the dynamics of active particles in complex environments see \cite{bdlrvv16}.

\begin{figure}[htbp]
\begin{center}
\includegraphics[width=.8\textwidth]{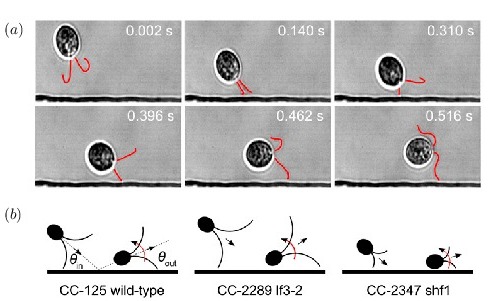}
\caption{(a) A \textit{Chlamydomonas reinhardtii} cell approaches a flat wall. Flagellar interactions with the surface lead to cell rotation until the cell swims back into the bulk at a geometry-specific departure angle. (b) The departure angle depends on the flagellar length, supporting the geometric view of the wall interaction. Reproduced from \cite{kdpg13} with permission.}
\label{Kantsler_clamyhit}
\end{center}
\end{figure}
Another recent experiment showed very clearly that the wall interaction may depend sensitively upon the microorganism geometry: Kantsler \textit{et al.}~found that {\it Chlamydomonas} algae cells scatter from a flat wall due to contact between its flagella and the surface (Fig.~\ref{Kantsler_clamyhit}a), so that the interaction is highly dependent on the cell body shape and flagellar lengths \cite{kdpg13}. With each stroke, one or both flagella are in contact with the surface and the cell body rotates as a consequence of the resultant torque. Rotation continues until there is no further flagellar contact with the surface, at which point the cell swims back into the bulk fluid. As reproduced in Fig.~\ref{Kantsler_clamyhit}b, the study included an examination of mutant cells with flagella of different length, and cells were found to depart from the surface on average at a critical reflection angle predicted by a simple geometrical consideration; the departure angle of wild-type {\it Chlamydomonas} was reported to have a narrow peak around $\theta_c=16^\circ$. The nature of the interaction suggests that the departure angle relative to the wall is independent of the angle of incidence, distinguishing this interaction from that of a classical billiard-like specular reflection. This angular rectification was also observed by Spagnolie \& Lauga in an analytical study of a model ``potential flow squirmer'' (a model microorganism with a prescribed tangential surface velocity) swimming near a surface \cite{sl12}.

Motivated by these behaviors of microorganisms and active particles near walls, we explore the billiard-like motion of a swimming body which, upon wall impact, swims away with a fixed angle of departure independent of the angle of incidence. While classical billiard dynamics --- in which a particle undergoes symmetric (specular) reflections with each wall collision --- is a well-developed problem in mathematics \cite{cm06}, the {\it microorganism billiard} motion of our investigation has not to our knowledge been studied even in simple domains. A similar type of non-specular reflection law also appears in spiral waves in bounded media~\cite{langham_non-specular_2013}, cavity solitons~\cite{prati_cavity_2011}, optical microcavities~\cite{altmann_non-hamiltonian_2008}, and even in internal gravity waves in stratified fluids \cite{ml95,mbsl97,maas2005wave}. Techniques used to study the classical billiard system, such as periodic reflections of the fundamental domain to tile a surface~\cite{Zorich2006}, hinge upon symmetry not present in our system.  In the mathematics literature, \emph{nonstandard reflection laws} have recently been studied~\cite{markarian_pinball_2010,del_magno_chaos_2012,del_magno_srb_2014}, but the focus has been mostly on contracting laws.  The \emph{slap map}~\cite{del_magno_srb_2014} applies to our setting for the very limited case of perpendicular reflection. 

We begin by studying two-dimensional microorganism billiard dynamics inside a regular polygon of arbitrary degree, where a one-dimensional return map is used to characterize the possible dynamics into stable periodic trajectories or unstable, chaotic orbits.  We show that the dynamics are robust to the introduction of weak random fluctuations. The Lyapunov exponent describing the dynamics of an ensemble of swimmer trajectories can be positive or negative, can have extremal values, and can have discontinuities depending on the degree of the polygon. Using the results for the stable periodic trajectories we propose a passive sorting device to trap swimmers of two different departure angles into two separate bins. Finally, a similar method of investigation is applied to the external problem, wherein the body swims through a periodic array of square obstacles. The results may contribute to our understanding of swimming microorganisms in porous environments, and suggest potential applications in entrapment and passive hydrodynamic sorting, and bioremediation.

The paper is organized as follows. In \S\ref{sec: internal} we study microorganism billiard dynamics inside a regular polygon, where consecutive and non-consecutive wall impacts are considered.  An explicit formula for the stable periodic trajectory is derived, and is shown to be robust to the addition of weak random fluctuations. The invariant measure and the Lyapunov exponent for the trajectories are also discussed. A sorting device is then proposed in \S\ref{sec: sorting}. Finally, in \S\ref{sec: external} we turn our attention to the external problem, billiard-like swimming in a periodic field of square obstacles, where lessons from the interior problem are instructive but the particle trajectories are more widely varied. We conclude with a discussion in \S\ref{sec: discussion}.

\begin{figure}[htbp]
\begin{center}
\includegraphics[width=.5\textwidth]{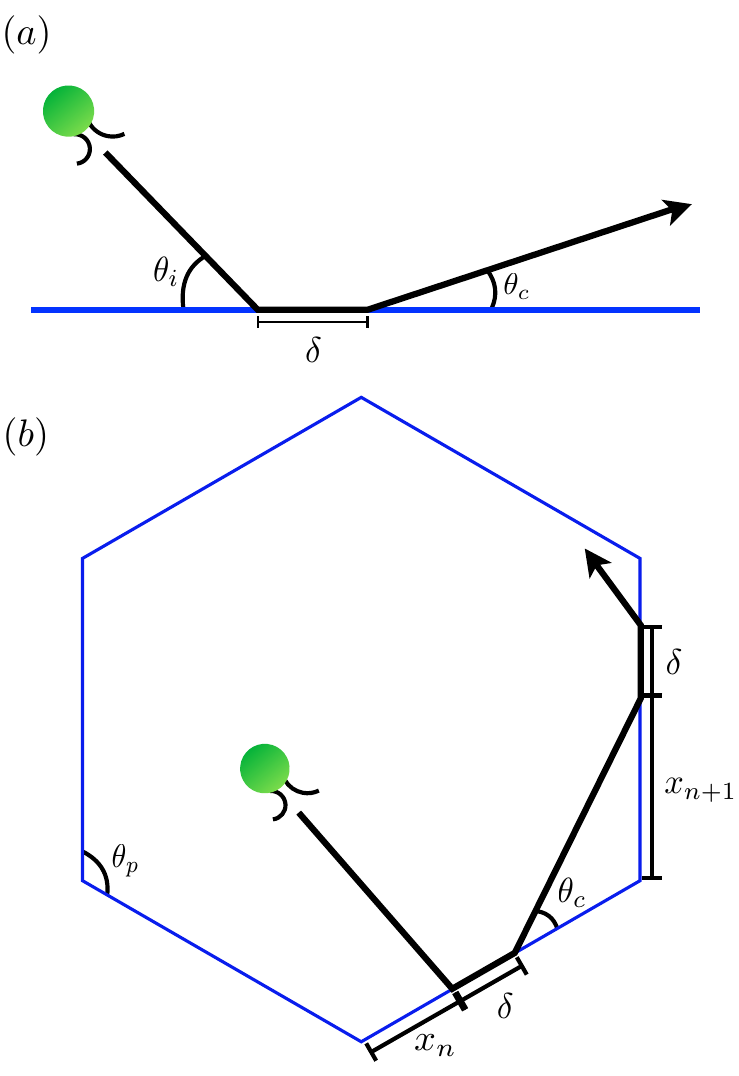}
\caption{(a) Schematic representation of microorganism billiard wall interactions. Regardless of the incident angle, $\theta_{i}$, the swimmer slides along the wall a distance $\delta$ and then departs with angle $\theta_{c}$. (b) Microorganism billiard inside a regular hexagon with interior angle $\theta_p=2\pi/3$. The $n$th point of impact, $x_{n}$, is measured from the vertex just behind the swimmer.}
\label{swimmer_hit_example}
\end{center}
\end{figure}

\section{Microorganism billiards inside a regular polygon}\label{sec: internal}

\subsection{Adjacent wall reflections lead to stable periodic orbits}\label{Internal1}

We begin by studying the deterministic microorganism billiard inside a regular polygon with $N$ sides, in the case where the body can only move from each wall to an adjacent wall. Figure~\ref{swimmer_hit_example}a illustrates the collision interaction with any wall; regardless of the angle of incidence, $\theta_i$, the body departs from the surface at a critical departure angle, $\theta_c\in(0,\pi/2)$. We include the possibility that the body translates a distance $\delta$ between impact and departure. The constraint that the swimmer only impacts an adjacent wall can be written in terms of the departure angle; namely, with the interior angle of the polygon given by $\theta_{p}=(N-2)\pi/N$, we require $\theta_c\in (0,\pi/N)$.

\begin{figure}[thbp]
\begin{center}
\includegraphics[width=.8\textwidth]{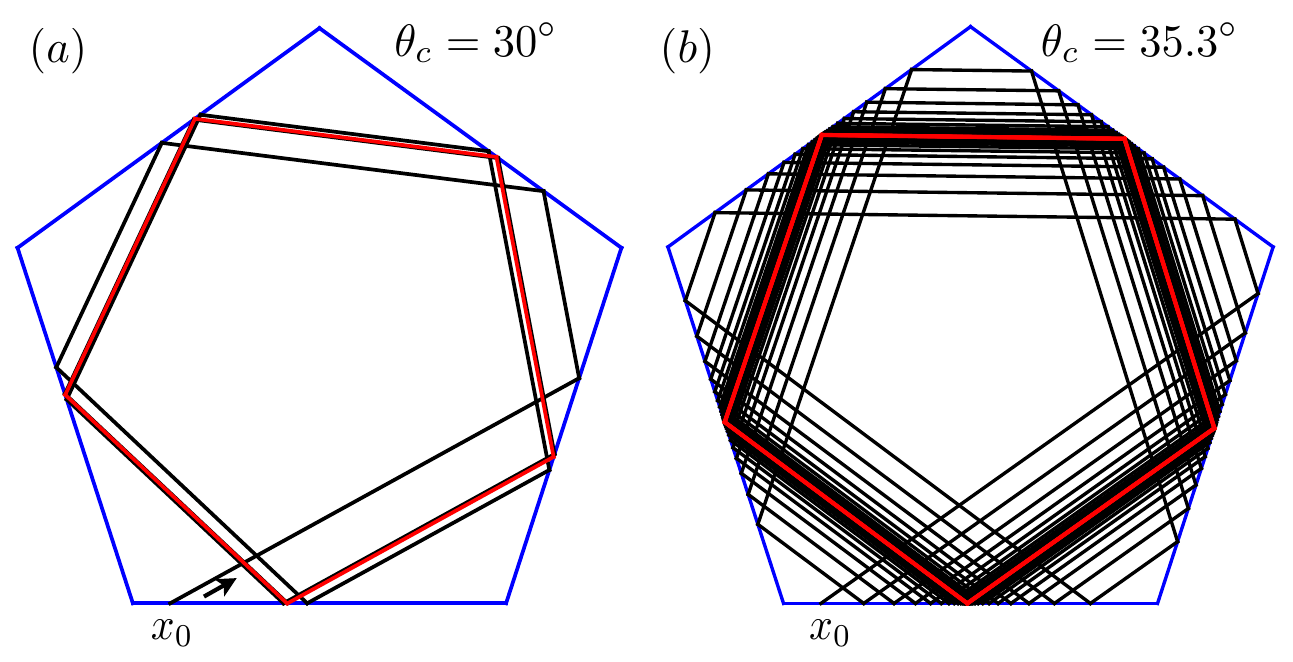}
\caption{(a) The trajectory with departure angle $\theta_c=30^\circ$ and initial point $x_0=0.1$ inside a pentagon rapidly converges to the stable periodic orbit. The trajectory is shown in red after the first $15$ reflections. (b) The trajectory with departure angle $\theta_c=35.3^\circ$ and initial point $x_0=0.1$ converges more slowly to the stable periodic orbit. The trajectory is shown in red after the first $150$ reflections.}
\label{StablePentagons}
\end{center}
\end{figure}

To understand the nature of the dynamics we take advantage of the symmetry in the system and study a map of $[0,1]$ to itself, writing $x_n$ for the $n$th point of impact, measured from the vertex just behind the swimming cell (see Fig.~\ref{swimmer_hit_example}b). (This is called the \emph{reduced billiard map}~\cite{del_magno_srb_2014}.) Given $x_n$, the subsequent point of impact is found geometrically. Assuming for the moment that $\delta=0$, we find
\begin{subequations}
\begin{gather}
x_{n+1}=f(x_n)=\beta(1-x_{n}),\label{no_skip_beta}\\
\beta=\sin(\theta_{c})/\sin(2\pi/N-\theta_{c})\label{no_skip_beta_2},
\end{gather}
\end{subequations}
where $\beta<1$ in the case of adjacent wall impacts. Writing $x_n=\beta(1-x_{n-1})$, and so on, results in an explicit expression for the $n$th position,
\begin{align}
x_{n}&=(-\beta)^n x_{0}-\sum\limits_{i=1}^n (-\beta)^i\nonumber\\
&=(-\beta)^n x_{0}+\beta\frac{1-(-\beta)^n}{1+\beta}.
\end{align}
Since $\beta<1$, the dynamics lose memory of the initial position $x_0$ exponentially fast in the number of impacts and the body settles to a periodic orbit with
\begin{gather}\label{adj_xstar}
x^*=\lim_{n\to \infty} x_n = \frac{\beta}{1+\beta}\, ,
\end{gather}
which is always less than $1/2$. Consider a uniform collection of swimmers departing from the entire length of one wall. By Eq.~\eqref{no_skip_beta}, the swimmers travel as a ``beam'' through the domain until colliding with the adjacent wall in a smaller region of length $\beta$, so that $\beta$ may be viewed as a stretch (or compression) factor. In the present setting the beam of swimmers is focused into a smaller region at every iteration, so that the trajectory rapidly converges to an asymptotically stable periodic orbit. Examples of this situation are shown in Fig.~\ref{StablePentagons}a\&b, with $\theta_c=30^\circ$ and $\theta_c=35.3^\circ$, both less than $36^\circ$ and so resulting solely in adjacent wall reflections. The stable periodic orbits are, as they must be, inscribed pentagons, and the rate of convergence depends on the stretch factor $\beta$. In the second case $\beta$ is very nearly one, and the convergence is slow.

\begin{figure*}[thbp]
\begin{center}
\includegraphics[width=.98\textwidth]{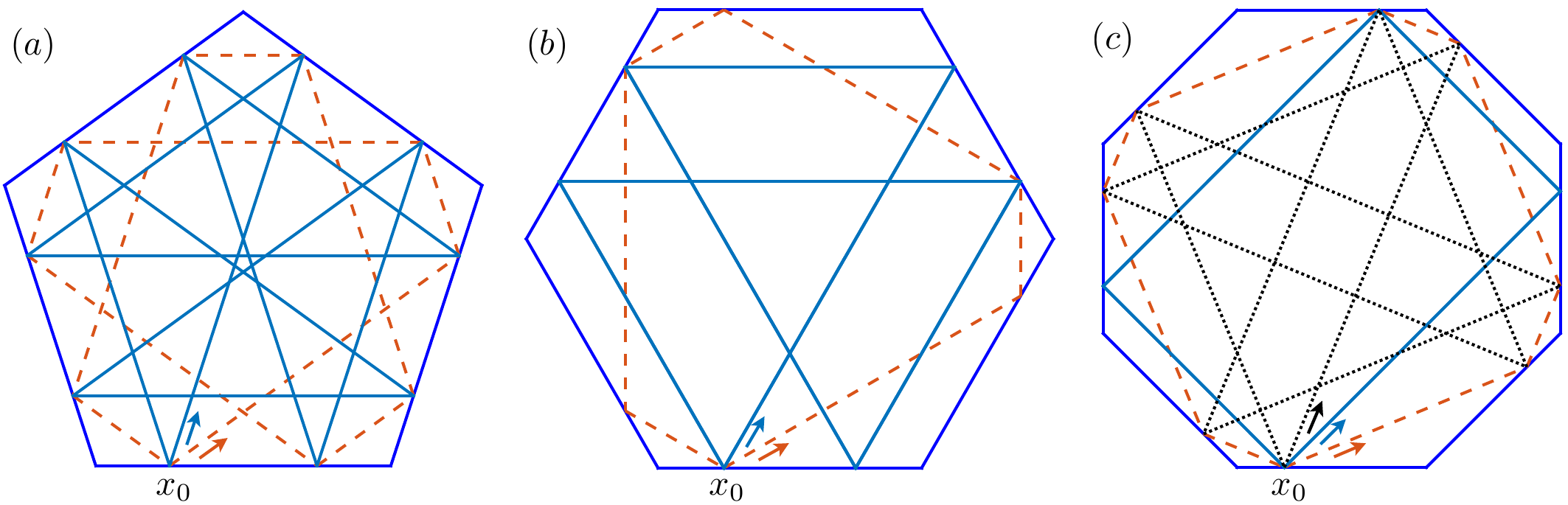}
\caption{When $\theta_c=k\pi/N$, each initial point $x_0$ is part of a neutrally stable periodic orbit. Shown are the (a) pentagonal, (b) hexagonal, and (c) octagonal boundary geometries, and trajectories for $k=1$ (dashed lines), $k=2$ (solid lines), and in the last case, $k=3$ (dotted line), all with $x_0=0.25$. Arrows indicate the swimming direction.}
\label{Periodic_568}
\end{center}
\end{figure*}

However, if we allow for the boundary case $\theta_c=\pi/N$, then the full length of the initial wall maps to the full length of the adjacent wall. Correspondingly we have $\beta=1$, and instead of the trajectory settling to one stable periodic orbit we now find that every initial point $x_0$ resides on a neutrally stable periodic orbit. For motion inside a square, for instance, the special angle is $\theta_c=\pi/4$, and any inscribed rectangle represents a neutrally stable trajectory. An example is shown for motion inside a pentagon with $x_0=0.1$ in Fig.~\ref{Periodic_568}a as a dashed line. Generally, an even number of walls leads to periodicity with a fundamental period (i.e., with $N-1$ reflections), while an odd number of walls leads to periodicity with twice the fundamental period (with $2N-1$ reflections). The initial position $x_0=1/2$ is a special case, for which the swimmer returns to the initial point after $N-1$ reflections for any $N$.

Returning to the case that $\delta \neq 0$, so that the body slides along the wall a fixed distance $\delta$ after each wall collision before departing, the mapping from the departure point $x_n$ to the next departure point $x_{n+1}$ is easily modified to $x_{n+1}=\beta(1-x_n)+\delta$ (assuming that $\delta$ is small enough that the swimmer does not slide past a vertex, $x_{n+1}<1$), resulting in the trajectory
\begin{align}\label{nth_hit_position}
x_{n}=(-\beta)^n x_{0}+(\beta+\delta)\frac{1-(-\beta)^n}{1+\beta}.
\end{align}
Hence, the motion still settles to an asymptotically stable periodic orbit in this case, with a shift in the fixed point of departure to $x^*=(\delta+\beta)/(1+\beta)$. More generally, the sliding distance $\delta$ might depend on the angle of incidence, as for model potential flow squirmers \cite{sl12}  and model ``pusher'' particles \cite{smbl15}. However, under the assumption of adjacent wall reflections, since the angle of incidence is independent of the departure position $x_n$, then $\delta$ is in fact constant, and the previous theory applies.

\subsection{Non-adjacent wall reflections: periodic and chaotic dynamics}\label{Internal2}

We will return to the role of random fluctuations in the dynamics, but first we will show that strong disorder in the system can be generated by the dynamical system alone once the constraint of purely adjacent wall reflections is relaxed. We now allow that the swimmer's next wall of impact may depend not only on $\theta_c$ but also on its position $x_n$. Fortunately, from the perspective of mathematical tractability, for a given departure angle a swimmer departing from any one wall can only possibly reach at most two other walls in the deterministic setting (which can be understood geometrically by inscribing isosceles trapezoids inside the regular polygon). For this study we set the sliding distance $\delta$ to zero, and we will only consider departure angles with $\theta_c<\pi/2$.

For one of a set of special departure angles, $\theta_c=\pi k/N$, where $k\in \{1,2,...,N-2\}$, the beam of swimmers collides with only one other wall. For $k\leq (N-1)/2$, the map is given trivially by $x_{n+1}=1-x_n$. Just as was seen for the adjacent wall case ($k=1$), there are an infinite number of neutrally stable periodic trajectories, since $x_{n+2}=x_n$. Let $p k= m N$, where $p$ and $m$ are the smallest integers such that this relation holds. Then any initial point $x_0$ corresponds to a neutrally stable orbit with a period (measured in the number of hits) equal to $p$ if $p$ is even or $x_0=1/2$, and period $2p$ if $p$ is odd and $x_0 \neq 1/2$.

Sample trajectories beginning at $x_0=0.25$ inside the $N=5$, $N=6$, and $N=8$ polygonal domains are shown in Fig.~\ref{Periodic_568}a-c, for $k=1$ (dashed line trajectories), $k=2$ (solid line trajectories), and in the last case, $k=3$ (the dotted line trajectory). For $N=5$, $p=5$ for both $k=1$ and $k=2$, so that a full period requires $10$ reflections. For $N=6$, $p=6$ for both $k=1$ and $k=2$, so that a full period requires $6$ reflections. Finally, for $N=8$, $p=8$ for $k=1$ and $k=3$, while $p=4$ for $k=2$. In this last case the neutrally stable periodic rectangle inscribed inside a square appears, as the dynamics do not explore the regions where the octagonal boundary differs from a square boundary, and we are back to the adjacent wall case.

\begin{figure}[thbp]
\begin{center}
\includegraphics[width=.86\textwidth]{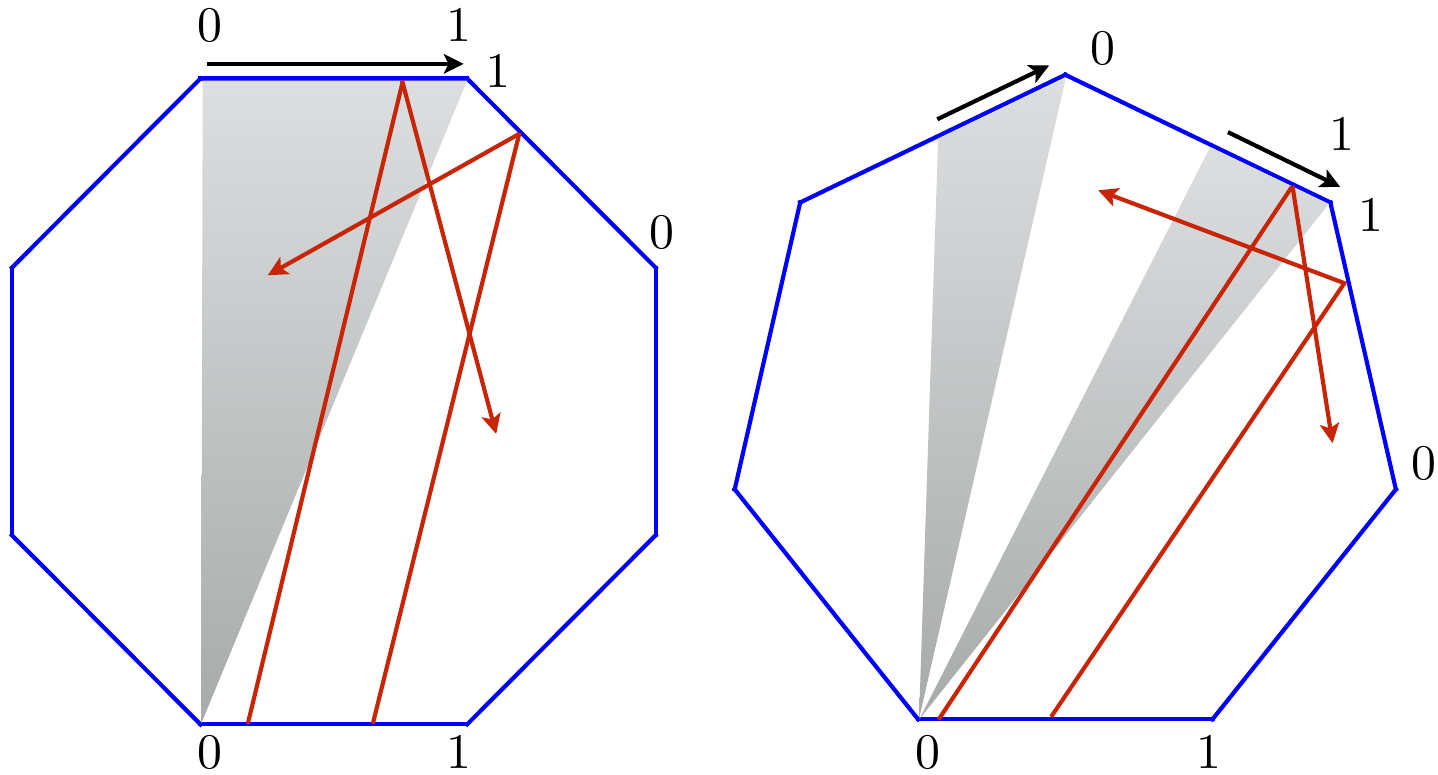}
\caption{(Left) When $N$ is even, the swimming direction at the next reflection requires an orientation reversal for the map when $\theta_c \in((N-2)\pi/2N,\pi/2)$, as indicated for the case $N=8$. Two sample trajectories leaving from the bottom wall with $\theta_c=76^\circ$ are shown in red. (Right) When $N$ is odd, a reversal is required when $\theta_c\in ((N-3)\pi/2N,(N-1)\pi/2N)$ and when $\theta_c\in((N-1)\pi/2N,\pi/2)$, as indicated for the case $N=7$. Two sample trajectories leaving from the bottom wall with $\theta_c=56^\circ$ are shown in red.}
\label{Flipping}
\end{center}
\end{figure}

Meanwhile, an additional subtlety must be taken into account for a small range of departure angles, specifically those for which the swimmer, as it approaches the next wall of impact, has changed its orientation relative to the mapping scheme: the direction of reflection must always be towards increasing~$x$.  When $N$ is even, the swimming direction at the next reflection requires an orientation reversal for the map when $\theta_c \in((N-2)\pi/2N,\pi/2)$. When $N$ is odd, a reversal is required when $\theta_c\in ((N-3)\pi/2N,(N-1)\pi/2N)$ and when $\theta_c\in((N-1)\pi/2N,\pi/2)$. Figure~\ref{Flipping} shows the cases $N=7$ and $N=8$. However, every wall may still be identified with every other wall if the direction of increasing $x$ is always the direction of swimming, also indicated in Fig.~\ref{Flipping}.

\begin{figure}[thbp]
\begin{center}
\includegraphics[width=.9\textwidth]{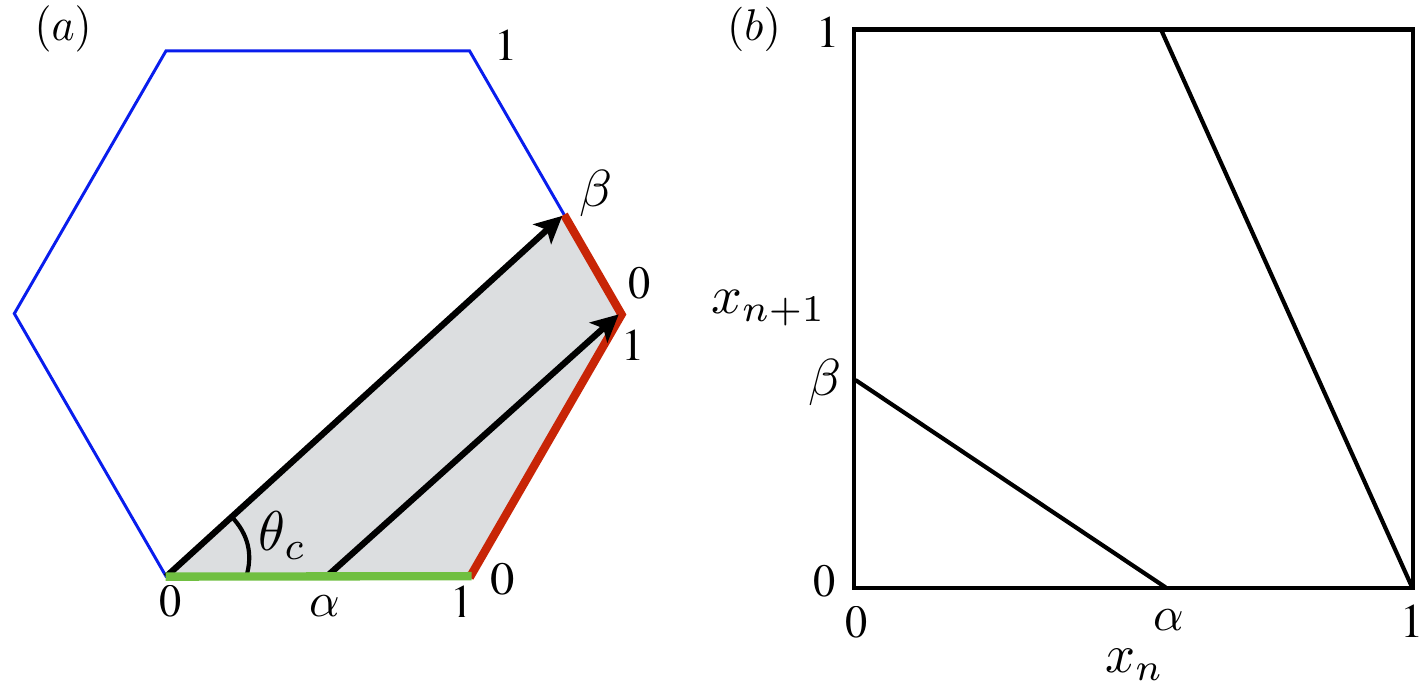}
\caption{(a) For motion inside a hexagon with $\theta_c\in(30^\circ,60^\circ)$, a swimmer departing from one wall may collide either with the adjacent wall or the subsequent wall. A ``beam'' of all possible swimmers leaving from the same wall is shaded grey. Swimmers departing from $x\in(\alpha,1)$ collide with the adjacent wall, an unstable, stretching region; swimmers departing from $x\in(0,\alpha)$ skip a wall and collide with a stable, focusing region.  (b) The piecewise linear map for $\theta_c=42^\circ$. The slope indicates beam stretching or focusing.}
\label{swimmer_beam_hexagon_skip}
\end{center}
\end{figure}

For departure angles in between these special values, a beam of swimmers extending from the full length of the original wall is bisected and collides with two adjacent walls. The map from $x_n$ to $x_{n+1}$ is piecewise linear with at most two distinct domains. In the mapping approach we lose information about the sequence of walls visited by the swimmer (though it can be deduced \textit{a posteriori}), but the reduction in dimensionality is crucial to understanding the dynamics. For example, consider a microorganism inside a hexagonal billiard, with interior angle $\theta_p=120^\circ$, and departure angle $\theta_c=42^\circ$, so that the swimmer may arrive next at either the adjacent wall or the subsequent wall, as illustrated in Fig.~\ref{swimmer_beam_hexagon_skip}a. The shaded region shows the space covered by the original beam of swimmers departing from the entire length of one wall. The map from $[0,1]$ to itself (see Fig.~\ref{swimmer_beam_hexagon_skip}a-b) takes $0$ to a point $\beta$, and a point $\alpha$ to $0$ (by convention due to the ambiguity at the vertex), and is given by
\begin{align}\label{tent_map_simplest_case}
 x_{n+1}= f(x_n)= \begin{cases}
 \displaystyle \beta\alpha^{-1}(\alpha-x_n)& x_n \leq \alpha ,\\
 \displaystyle (1-\alpha)^{-1}(1-x_n) & x_n > \alpha,
\end{cases}
\end{align}
where $(\alpha,\beta)\approx(0.54,0.37)$. The definition of $\beta$ is just as in Eq.~\eqref{no_skip_beta_2}, where it also corresponds to the image of the point $x=0$. The values $\alpha$ and $\beta$ for arbitrary $(N,\theta_c)$ are included in ~\ref{apx}. Using the beam analogy, the region $x\in(\alpha,1)$ maps to the full length of the adjacent wall, corresponding to a decreased swimmer density upon arrival and a slope of the map with absolute value greater than one. Meanwhile, the region $x\in (0,\alpha)$ maps to a region of length smaller than $\alpha$, corresponding to an increase in the swimmer density and a slope of the map with absolute value less than one.

\begin{figure}[thbp]
\begin{center}
\includegraphics[width=.9\textwidth]{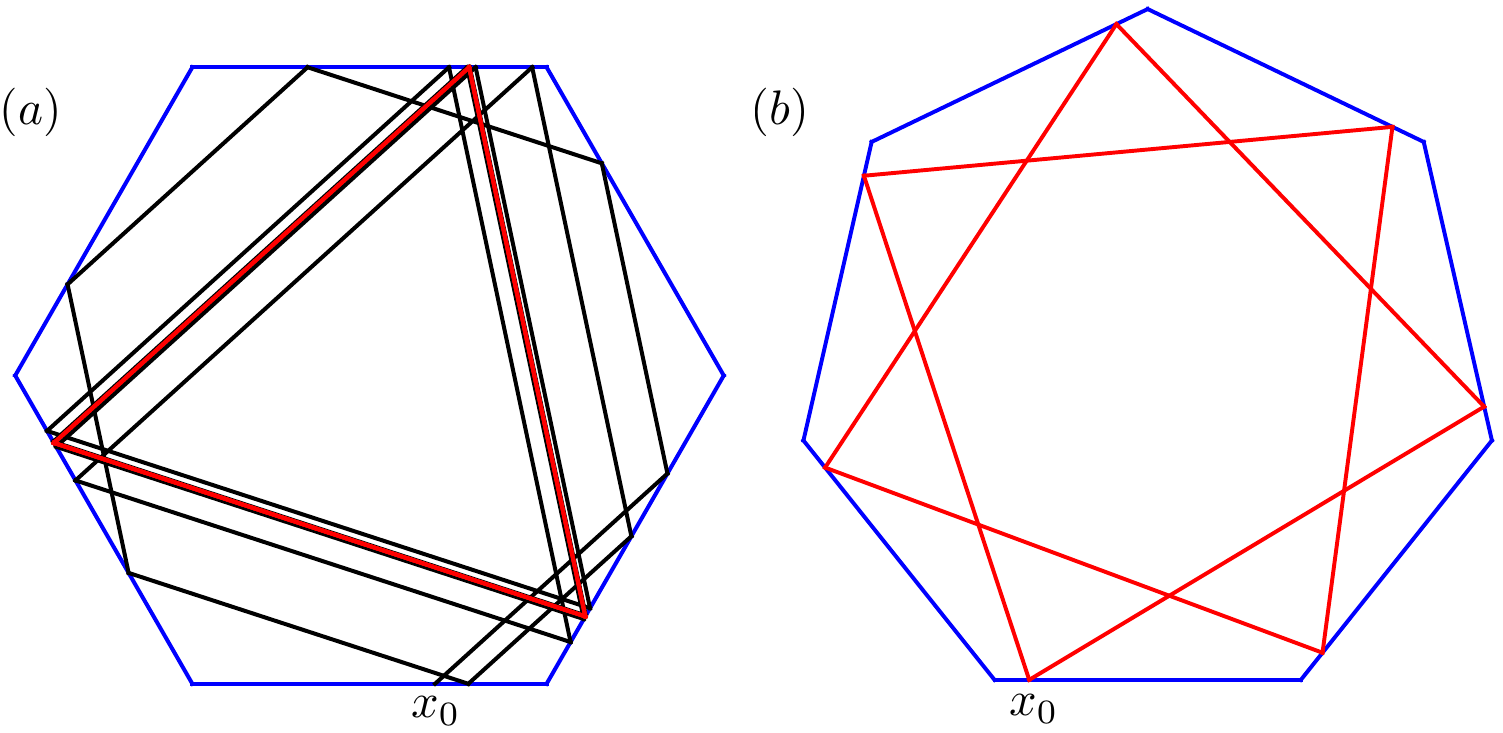}
\caption{(a) The trajectory of a swimmer with departure angle $\theta_c=42^\circ$ and initial position $x_0=0.685$ in a hexagon ($N=6$). After seven collisions with walls adjacent to each wall of departure the swimmer skips a wall and settles rapidly to a consistently wall-skipping periodic triangular orbit. The trajectory is colored red after the first 30 reflections. (b) A stable, wall-skipping periodic orbit with a prime number of walls cannot trace out a regular polygon (here $N=7$, $\theta_c=30^\circ$, and $x_0=x^*=0.093$).}
\label{NewHex}
\end{center}
\end{figure}

The analogy illuminates individual swimmer dynamics: regions associated with a decreasing swimmer density, or stretching, are unstable for the swimmer trajectory, regions where the density of swimmers is unchanged are neutral, and regions of increasing density, or focusing, are stable. For example, the trajectory with departure angle $\theta_c=42^\circ$ and initial point $x_0=0.685$ inside a hexagon is shown in Fig.~\ref{NewHex}a. The swimmer begins by reflecting seven times with walls adjacent to each wall of departure. These reflections take place in the unstable region of the map, and the swimmer is eventually expelled to the other (stable) region where it rapidly settles to a consistently wall-skipping periodic triangular orbit. Note that this stable trajectory is precisely the one predicted for motion inside a regular triangle; the swimmer only visits three walls and cannot distinguish the hexagonal geometry from the triangular geometry. The stretch factor $\beta\alpha^{-1}$ is less than one, resulting in stable periodic orbits. If the number of walls is prime, however, a stable wall-skipping orbit cannot trace out a regular polygon, and generally results in a more complex trajectory as observed in Fig.~\ref{Periodic_568}a (solid line) for $N=5$ or Fig.~\ref{NewHex}b for $N=7$.

\begin{figure}[thbp]
\begin{center}
\includegraphics[width=.96\textwidth]{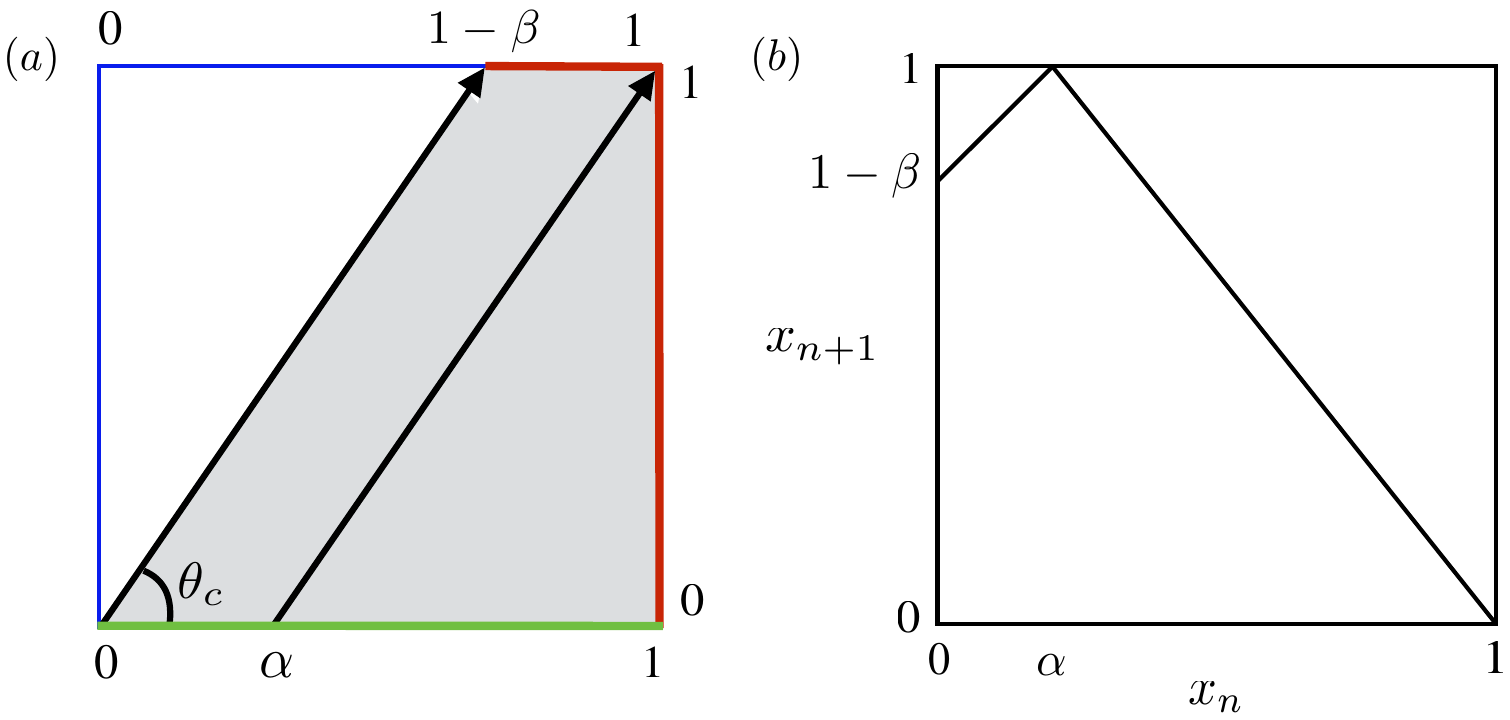}
\caption{(a) Motion inside a square with $\theta_c\in(\pi/4,\pi/2)$ is illustrated. Unlike in Fig.~\ref{swimmer_beam_hexagon_skip}, only neutral and unstable regions exist, and the resulting dynamics are chaotic. (b) An orientation reversal results in a continuous map from $[0,1]$ to itself.}
\label{swimmer_square_map_beam}
\end{center}
\end{figure}

\begin{figure*}[htbp]
\begin{center}
\includegraphics[width=.9\textwidth]{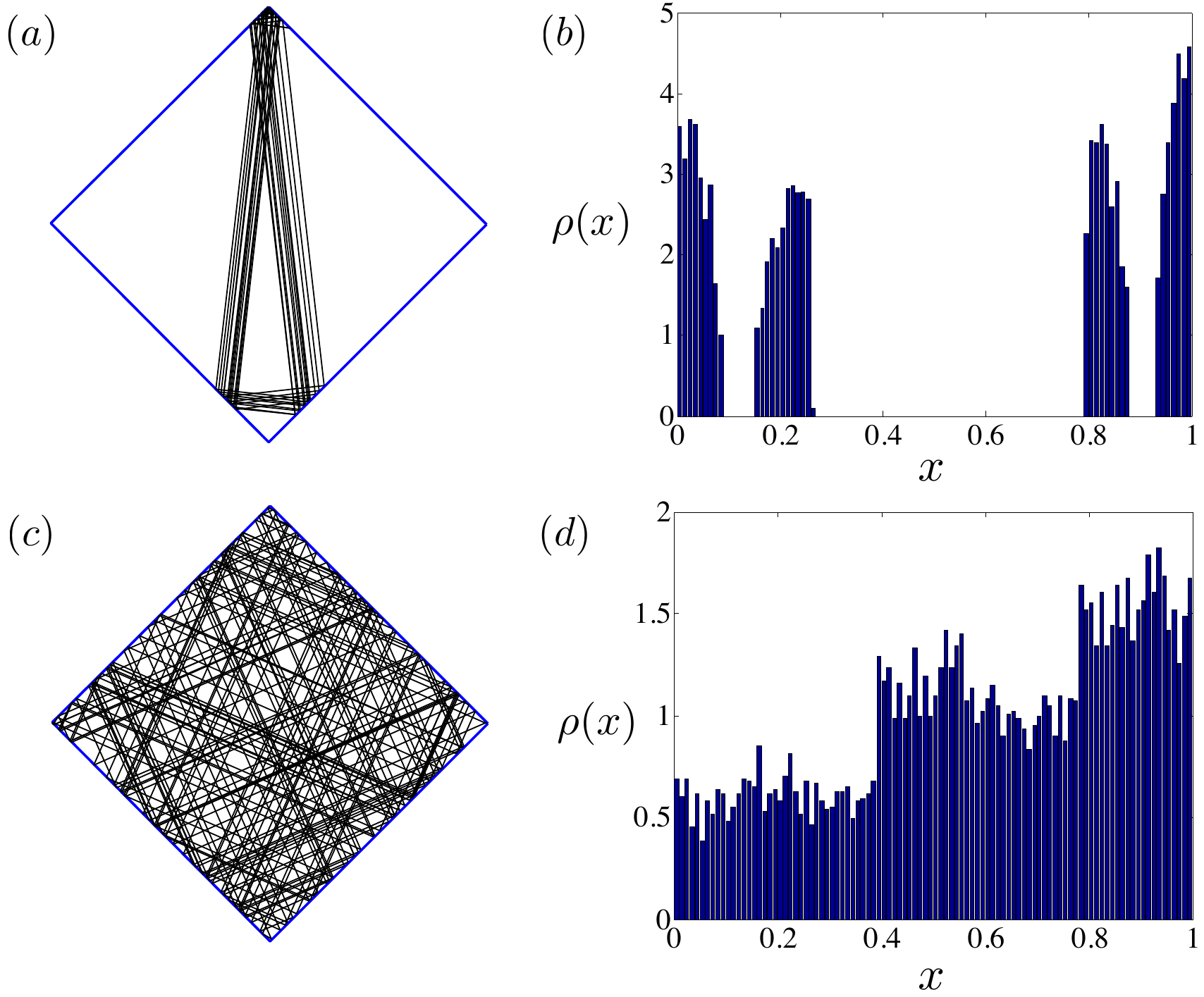}
\caption{(a) Simulation of a swimmer in a square with $\theta_{c}=52^{\circ}$. (b) Simulated prediction of the invariant measure for $\theta_{c}=52^{\circ}$. The final distribution shows that there are four domains that have nearly equal numbers of swimmers. Swimmers that start initially close together may end up in drastically different places with in these domains or possibly in entirely separate domains which makes this system chaotic. (c) Simulation of a swimmer in a square with $\theta_{c} = 72^{\circ}$. (d) Simulated prediction of the invariant measure for $\theta_{c} = 72^{\circ}$. The system appears nonuniform and depicts a seemingly ergodic system.}
\label{invMeasPaper}
\end{center}
\end{figure*}

Consider now a seemingly simpler example, microorganism billiards inside a square with $\theta_c\in(\pi/4,\pi/2)$, as illustrated in Fig.~\ref{swimmer_square_map_beam}a. When the swimmer approaches the wall opposite the wall of departure it has changed its orientation relative to the mapping scheme, and the direction of increasing $x$ must be reversed. The piecewise linear map for this case is given by
\begin{align}\label{tent_map_simplest_case2}
 x_{n+1}=\begin{cases}
 \displaystyle 1-\beta\alpha^{-1}(\alpha-x_n)& x_n \leq \alpha ,\\
 \displaystyle (1-\alpha)^{-1}(1-x_n) & x_n > \alpha,
\end{cases}
\end{align}
which is now a continuous map due to the orientation reversal on the opposite wall (see Fig.~\ref{swimmer_square_map_beam}b). There remains one large unstable, stretching region corresponding to adjacent wall collisions, but now the opposite wall is a neutral region in which the density of swimmers in an incident beam is unchanged. The long-term dynamics of an individual swimmer are determined by the interplay between the unstable and neutral regions, and in particular the frequency of visits to the unstable stretching region and the degree of stretching there. Two such trajectories are shown with departure angles $\theta_c=52^\circ$ and $\theta_c=72^\circ$ in Fig.~\ref{invMeasPaper}a\&c. The dynamics are chaotic in both examples, but there is a clear distinction between the resulting trajectories. For $\theta_c=52^\circ$, after a few transient reflections (not shown) the trajectory is confined to a small region of the interior domain, while in the second case the dynamics do not at first glance appear to be contained in any such subregion. To study such long term dynamics in more detail we will proceed in the next section to explore the invariant measure and the Lyapunov exponent.

\begin{figure}[htbp]
\begin{center}
\includegraphics[width=.86\textwidth]{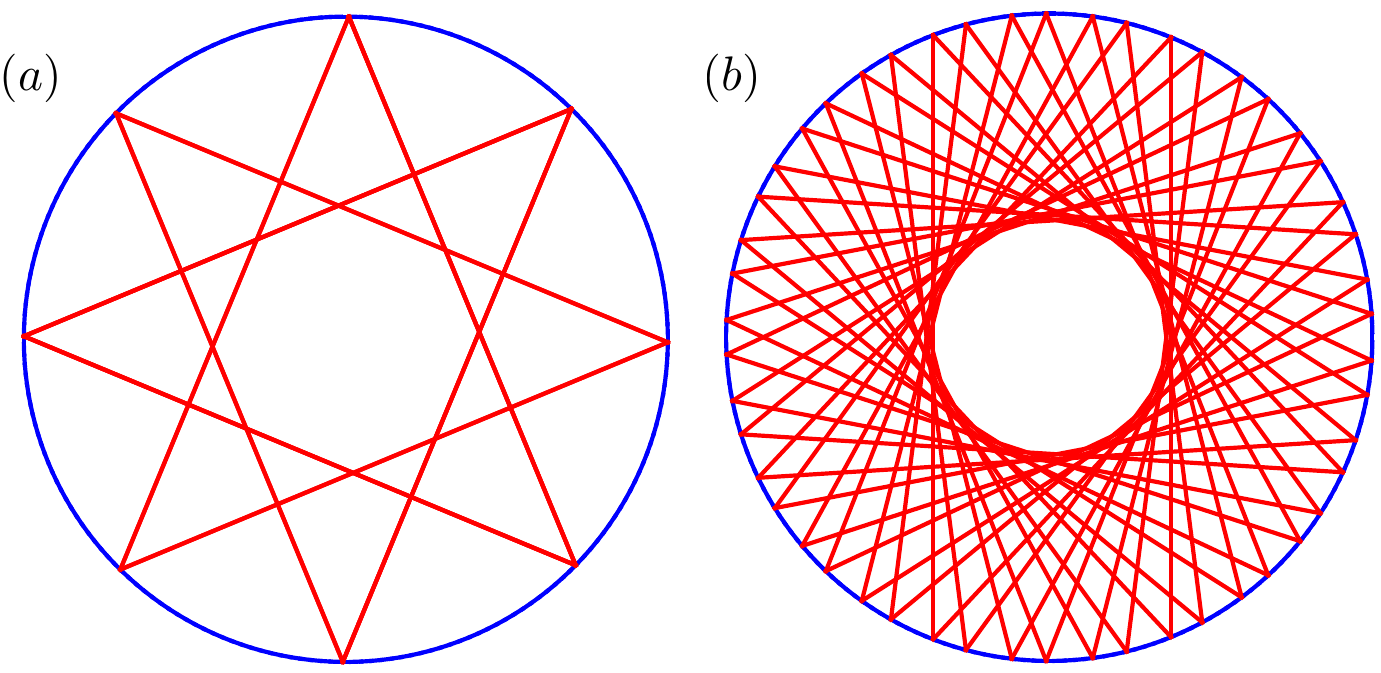}
\caption{With $N=200$ sides, and initial point $x_0=0.2$, periodic trajectories are found for the special angles: (a) $\theta_c=75\pi/200$ (period $p=8$), and (b) $\theta_c=76\pi/200$ (period $p=50$).}
\label{Circles}
\end{center}
\end{figure}
For large, finite values of $N$, integer multiples of $\pi/N$ are special values of~$\theta_c$ for which the beam collides with only one other wall and all initial points $x_0$ give a neutrally stable periodic orbit. Figure~\ref{Circles}a shows such a finite estimate of such an example, with $N=200$, $\theta_c=75\pi/200$, and $x_0=0.2$. If the beam intersects two walls, let $\theta_c=(a+b)^{-1}\left[a(k\pi/N)+b(k+1)\pi/N\right]$ for some $a,b>0$, so that $k=N\theta_c/\pi-b(a+b)^{-1}$. Then, for large $N$,
\begin{gather}
(\alpha,\beta)\sim\frac{b}{a+b}(1,1)+\frac{\pi a b}{(a+b)^2\tan(\theta_c)N}(1,-1),
\end{gather}
so that
\begin{gather}
\beta\alpha^{-1}\sim 1-\frac{2\pi a}{(a+b)\tan(\theta_c)N}<1.
\end{gather}
Hence, for large but finite $N$ there remains a stable focusing region, but the period of the stable periodic orbit can be quite large. Figure~\ref{Circles}b shows a trajectory with $N=200$, $\theta_c=76\pi/200$, and $x_0=0.2$, where the period is found to be $p=50$ (the smallest $p$ for which $76p=200m$ with $m,p$ integers).

Finally, in the limit as $N\to\infty$ (a circular boundary) the entire boundary may be mapped onto itself via the internal angle, $[0,2\pi]\to [0,2\pi]$ by writing $\theta_{n+1}=\theta_n+2\theta_c$ (assuming $\theta_c<\pi/2$). Periodic trajectories are therefore found whenever $\theta_c=(q/p)\pi$ for integer $q$ and $p$, and then $p$ is the period of the trajectory; otherwise, the map is chaotic but returns to within an arbitrarily small distance of any point on the circle infinitely often.

\subsection{Lyapunov exponent for the dynamics}\label{sec: Lyapunov}

A useful means of characterizing the long-term behavior of the billiard system is to study the nature of the invariant measure for the dynamics, or the distribution of swimmers on each wall that is preserved with each iteration of the mapping. The invariant measure for the case of purely adjacent wall reflections (see \S\ref{Internal1}) is simply the fixed point of the map, $x^*=f(x^*)$ in Eq.~\eqref{no_skip_beta}. Writing the normalized invariant distribution as $\rho(x)$, this case corresponds to a density with support at only one point, $\rho(x)=\delta_{x^*}(x)$, where $\delta_{x^*}(x)$ is a Dirac delta function centered at $x^*$.

The invariant measure for more irregular cases, such as for swimming in a square with $\theta_c\in(\pi/4,\pi/2)$, is more complicated but speaks to the different possible structures observed in the long-time behavior of the system, for instance whether or not the trajectories are ergodic over the polygon's interior (space-filling). The invariant measures for the cases $\theta_c=52^\circ$ and $\theta_c=72^\circ$, depicted in Fig.~\ref{invMeasPaper}b,d, were computed by taking a uniform distribution of initial positions, $x_0\sim U[0,1]$, and measuring their locations after $4000$ wall reflections (and finally, normalizing). In the first case, each individual swimmer is attracted to a small region of space, reflecting off of only four separate small domains. Depending on the initial position, the invariant set is one of four symmetric rotations of the region shown in Fig.~\ref{invMeasPaper}a. In the second case a single swimmer trajectory appears to cover the entire domain, as for a classical ergodic system.

It is natural to ask about the rate at which two neighboring trajectories converge together or diverge from each other --- the maximal Lyapunov exponent for the dynamical system.  This is defined as
\begin{align}\label{max_lyap_formula}
\lambda = \lim_{n\rightarrow\infty}\frac{1}{n}\sum_{i=0}^{n-1}\log|f'(x_i)|,
\end{align}
where $x_{i+1}=f(x_i)$. The distance along a wall between two swimmers after $n$ reflections is approximately $\Delta x_n\approx \Delta x_0 \exp(\lambda n)$, where~$\Delta x_0$ is their initial separation. For $\theta_{c}=52^{\circ}$, the Lyapunov exponent is approximately $\lambda=0.12$, while for $\theta_{c}=72^{\circ}$ the value is larger, $\lambda=0.47$. From these two numbers, we can determine that the configuration in Fig.~\ref{invMeasPaper}a undergoes approximately a third as much stretching as that in Fig.~\ref{invMeasPaper}c. This tells us one of two things about the system with $\theta_{c}=52^{\circ}$: either this system spends less time in the chaotic region or it has a chaotic region that does only a small amount of stretching per iteration.

\begin{figure*}[htbp]
\begin{center}
\includegraphics[width=.98\textwidth]{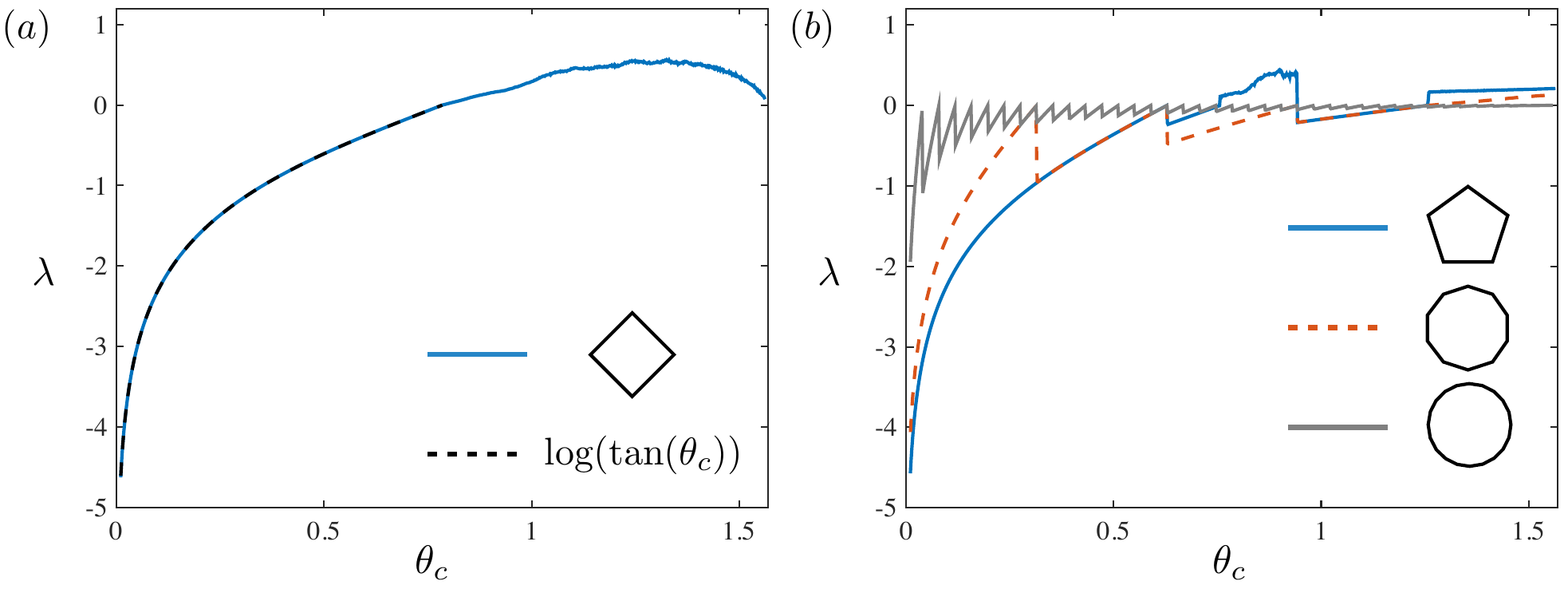}
\caption{(a) In a square domain, the maximal Lyapunov exponent is exactly $\lambda = \log(\tan(\theta_c))$ for $\theta_c<\pi/4$, and then increases to a very noisy peak region before decaying again, always bounded above by $1$. (b) The Lyapunov exponent is shown for trajectories internal to regular polygons with $N=5,10$ and $20$ sides. Regions of overlap are departure angles for which the swimmer cannot distinguish between geometries.}
\label{LyapunovCurves}
\end{center}
\end{figure*}

Figure \ref{LyapunovCurves}a shows the computed Lyapunov exponent as a function of $\theta_c$ for the square domain. For $\theta_c<\pi/4$, the dynamics are confined to adjacent wall reflections, and from Eq.~\eqref{no_skip_beta} we have that $|f'(x_i)|=\tan(\theta_{c})$. The stability of the dynamics in the adjacent wall case corresponds to negative values of $\lambda$, and specifically $\lambda = \log(\beta)=\log(\tan(\theta_{c}))$ is exact for $\theta_c<\pi/4$. For $\theta_c\in(\pi/4,\pi/2)$, however, the contribution to $\lambda$ depends on the frequency of reflections from the unstable region (where $|f'(x_i)|>1$) and the neutral region (where $|f'(x_i)|=1$). The Lyapunov exponent is very rough in the region where it is largest.

Figure \ref{LyapunovCurves}b shows the computed Lyapunov exponents for three other polygonal domains, with $N=5$, $10$, and $20$. When $\theta_c<\pi/N$ the swimmer only moves to the adjacent wall and $\lambda =\log(\beta)=\log(\sin(\theta_c)/\sin(2\pi/N-\theta_c))<0$, exactly. For one of the special angles $\theta_c=k\pi/N$ with integer $k$ and $\theta_c<\pi/2$, such that the departure wall maps perfectly onto one other wall, the periodic trajectory is neutrally stable and we have exactly $\lambda = 0$. For other departure angles, the dynamics can be stable and periodic, or unstable and chaotic. Regions of overlap are departure angles for which the swimmer cannot distinguish between the geometries, as in the octagon/square example in Fig.~\ref{Periodic_568} and the hexagon/triangle example in Fig.~\ref{NewHex}. For microorganism billiards inside a circle, the dynamics are always neutral, and the limiting behavior is $\lambda\to 0$ for all fixed $\theta_c$.

Note that the computed value $\lambda$ is the Lyapunov exponent of the one-dimensional map, which differs from a ``true'' Lyapunov exponent since the transit time between reflections varies throughout the dynamics. Defining the~$n$th transit time between reflections as ~$t_n$, let the total time $T_n = \sum_{i=0}^{n-1} t_i$, with~$t_i \le \tau$.  Here~$\tau$ is the maximum time between collisions, which depends on the diameter of the domain and the speed of the organism ($\tau = \text{diam}/U$).  The true Lyapunov exponent~$\Lambda$ (with units of inverse time) is
\begin{equation*}
\Lambda = \lim_{n\rightarrow\infty}\frac{1}{T_n}
\sum_{i=0}^{n-1}\log|f'(x_i)|,
\end{equation*}
which differs from~$\lambda$ (see Eq.~\eqref{max_lyap_formula}) by the factor of~$T_n$ rather than~$n$ in the denominator.  We thus have
\begin{equation}
  \Lambda = \lim_{n\rightarrow\infty}\frac{n}{T_n}\,\lambda_n,
\end{equation}
where~$\lambda_n$ is Eq.~\eqref{max_lyap_formula} without the limit.  By Oseledec's multiplicative ergodic theorem~\cite{Oseledec1968}, both limits defining~$\Lambda$ and~$\lambda$ must exist, which implies that the average time between encounters~$\overline{T} = \lim_{n\rightarrow\infty}T_n/n \le \tau$ must also exist.  Hence, we have
\begin{equation*}
  \Lambda = \lambda\,/\,\overline{T},\qquad
  \lvert\Lambda\rvert \ge \lvert\lambda\rvert/\tau.
\end{equation*}
Since $\Lambda$ and~$\lambda$ have the same sign, chaos in the map is reflected as chaos in the full system, as is the absence of chaos.

\subsection{Robustness to random fluctuations}\label{random_fluctuations}

The swimming trajectory of a microorganism is unlikely to be straight even over short distances.  This can be due to internal biological mechanisms, such as the run-and-tumble flagellar dynamics of {\it E.~coli}, thermal fluctuations, or other hydrodynamic effects.  Alternatively, the angle of departure is likely to vary somewhat with each wall interaction, which was found to be true for {\it Chlamydomonas} wall reflections by Kantsler \textit{et al.}~\cite{kdpg13}. We would like to know whether the stable periodic orbits found for the adjacent wall case, with $\beta<1$, survive the introduction of weak random fluctuations.

As a first approximation to the effect of randomness we assume a Gaussian distribution around the deterministic point of arrival, writing for adjacent wall interactions in a regular polygon,
\begin{gather}
x_{n+1}=\beta(1-x_{n})+\sigma Z_{n+1},
\label{rand}
\end{gather}
where $\beta$ is given in Eq.~\eqref{no_skip_beta}, $Z_n\sim N(0,1)$ are independent normal Gaussian random variables, and $\sigma$ is the standard deviation.  The solution to~\eqref{rand} is
\begin{align}
x_{n}=&(-\beta)^n x_{0}\nonumber\\
&+\sum\limits_{i=1}^{n-1} (-1)^{i+1}\beta^i(1-\sigma Z_{n-i})+\sigma Z_n\nonumber\\
=&(-\beta)^n x_{0}+\beta\frac{1-(-\beta)^n}{1+\beta}+\sigma\sum\limits_{i=0}^{n-1} (-\beta)^{i}\, Z_{n-i}.
\end{align}
The last term is a sum of normal random variables with different variances, which leads to
\begin{gather}
x_{n}=(-\beta)^nx_{0}+\beta\frac{1-(-\beta)^n}{1+\beta}+ \sigma\sqrt{\frac{ 1-\beta^{2n}}{1-\beta^2}}\,\tilde{Z}_n,
\end{gather}
where $\tilde{Z}_n\sim N(0,1)$. In the case $\beta<1$, the transient dynamics decay exponentially fast and the trajectory is simply given by
\begin{gather}
x_{n}\sim\frac{\beta}{1+\beta}+\frac{ \sigma \tilde{Z}_n}{\sqrt{1-\beta^2}},
\qquad n\gg 1.
\end{gather}
The dynamics are still focused around the stable limit point in the deterministic case, but with a Gaussian spread that retains a memory of the recent past. When $\beta\approx 1$ the variance of the dynamics may be very large, but this is also the setting where the other walls should really be taken into consideration. Nevertheless, for reasonably small $\sigma$ and for adjacent wall reflections, the dynamics are robust, settling neatly into the deterministic periodic orbit with only small deviations.

Another biologically relevant form of randomness, observed in the reflections of {\it Chlamydomonas} cells \cite{kdpg13}, is a random departure angle. If the $n$th departure angle is given by $\theta_c + \sigma Z_n$, again with $Z_n\sim N(0,1)$, then this results in a modified and random stretch factor $\beta\to \beta_n=\sin(\theta_c+\sigma Z_n)/\sin(\theta_p+\theta_c+\sigma Z_n)$. While large tail events may lead to a non-adjacent wall reflection, which could result in a significant shift of the trajectory, we will neglect such events for the present calculation. The mapping is then given by $x_{n+1}=\beta_n(1-x_{n})$, resulting in a trajectory
\begin{gather}
x_n =Q_{n}x_0-\sum_{j=1}^{n} Q_{j},\\
Q_j= (-1)^{j}\prod_{i=0}^{j-1} \beta_{n-1-i},
\end{gather}
which may be used for computational purposes but is not analytically tractable. If, however, we assume a small variance and Taylor expand about small $\sigma$, then $\beta_n\approx \beta+r \sigma Z_n$, where $r=\sin(\theta_p)/\sin^2(\theta_c+\theta_p)$, and then inserting into the above and neglecting all terms of $O(\sigma^2)$ we find
\begin{align*}
Q_j&=(-1)^{j}\prod_{i=0}^{j-1} (\beta+r \sigma Z_{n-1-i})\\
&\approx(-1)^{j} \left(\beta^j+r \beta^{j-1}\sigma \sum_{i=0}^{j-1}Z_{n-1-i}\right)\\
&=(-1)^{j} \left(\beta^j+ \beta^{j-1}r\sigma \sqrt{j}\tilde{Z}_j\right),
\end{align*}
where $\tilde{Z}_j\sim N(0,1)$.
Then, after summing over normal variables as before, we have
\begin{multline*}
x_n \approx (-1)^{n} \left(\beta^n+r \beta^{n-1}\sigma \sqrt{n}\tilde{Z}_n\right) x_0+\frac{\beta\left(1-(-1)^n\right)}{1+\beta}\\
+r \sigma \sqrt{\frac{1-\left(1+n \left(1-\beta^2\right)\right) \beta^{2 n}}{\left(1-\beta^2\right)^2}}\,\tilde{Z}'_n,
\end{multline*}
where $\tilde{Z}'_n\sim N(0,1)$. For $\sigma=0$ we recover the deterministic trajectory from Eq.~\eqref{nth_hit_position}, and for $n$ large the trajectory is again drawn to the deterministic fixed point, with
\begin{gather}
x_n\sim\frac{\beta}{1+\beta}+ \frac{r \sigma\tilde{Z}'_n}{1-\beta^2},
\qquad n \gg 1.
\end{gather}
Note that $r(1-\beta^2)^{-1}=\csc(\theta_p+2\theta_c)$, with $\theta_p=(N-2)\pi/N$ the interior angle of the polygon.

\section{Passive sorting with boundary geometry}\label{sec: sorting}

\begin{figure}[htbp]
\begin{center}
\includegraphics[width=.6\textwidth]{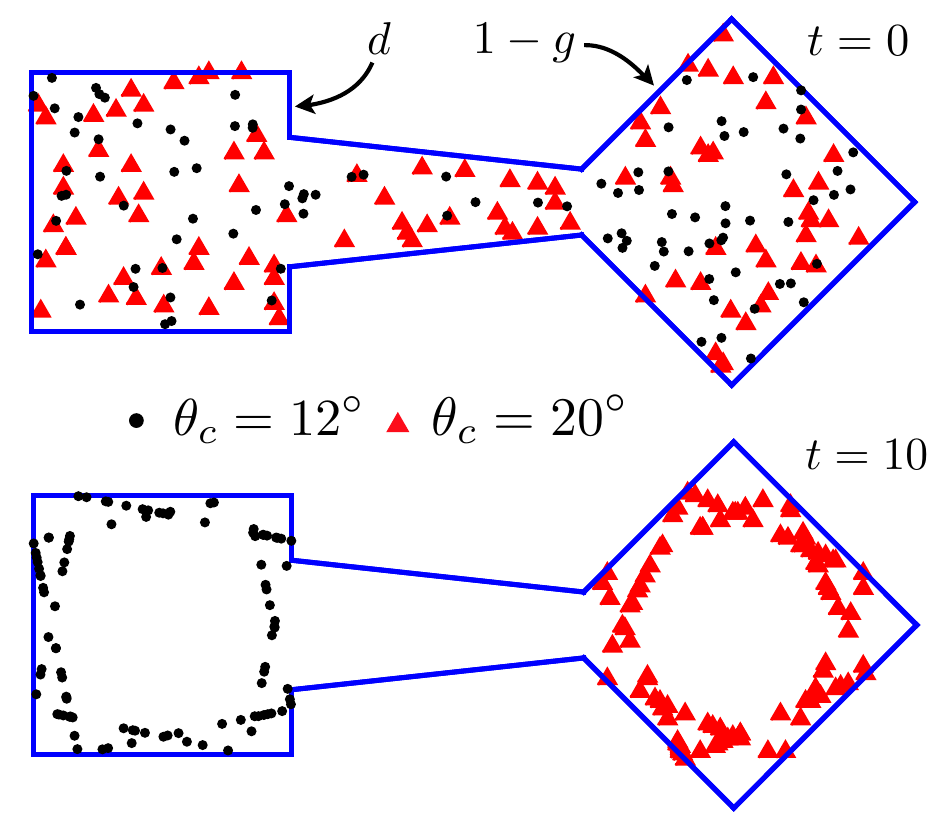}
\caption{Two hundred swimmers with unit speed, half with $\theta_c=12^\circ$ (circles) and half with $\theta_c=20^\circ$ (triangles), placed at random inside a simple sorting device. Their positions are shown at $t=0$ (top), and then again perfectly sorted at $t=10$ (bottom).}
\label{PerfectSorting}
\end{center}
\end{figure}

With an eye towards the manipulation of microorganism populations for basic biological research and potential engineering applications \cite{hd14}, numerous capture and sorting techniques have recently been designed using wedge-shaped boundaries \cite{kwl12,kpwl13,gjbmcmgs14}, chevron and heart-shaped chips \cite{rsmss14}, smooth microchannels \cite{mdgkd09,ceagr14,pdla14}, corrugated microchannels \cite{bqyfz13}, non-convex boundary geometries \cite{fbh14}, and regular patterned arrays \cite{vbvkb11,phs13,rr13}. Using the findings of the previous section we propose a new passive sorting technique for species with different departure angles. In particular, we consider swimmers with departure angles $\theta_c=12^\circ$ and $\theta_c=20^\circ$, which are the mean values for two different {\it Chlamydomonas} strains \cite{kdpg13}. The sorting device is shown in Fig.~\ref{PerfectSorting}; a channel connects two square chambers of unit area which are relatively rotated by $90^\circ$, and for the lengths $d$ and $g$ shown in the figure we choose $d=0.25$ and $g=0.18$. The centers of the squares are separated by a distance $(3+\sqrt{2})/2$. Initially, 100 of each swimmer type are distributed with random initial position and orientation in the device, but at $t=10$ (with swimmers moving at unit speed) the two strains are perfectly sorted.

The two swimmer types are separated by choosing the lengths $d$ and $g$ so that the fixed points of the map associated with each swimmer lie on a wall in one square and on a gap in the other. For a square domain, the fixed point is given by $x^*(\theta_c)=\tan(\theta_c)/(1+\tan(\theta_c))$ (see \S\ref{Internal1}). The two swimmers above are therefore successfully sorted when $x^*(12^\circ)<d,g < x^*(20^\circ)$, or $0.175<d,g<0.267$. The distance between the channels is not too important but without at least a small separation there can be more complicated effects near the opening in the leftmost square. If instead a rectangular channel (parallel walls) connects the square domains, it can be shown that the device sorts two swimmers so long as one has $\theta_c<22.5^\circ$ (exactly) and the other has $\theta_c>22.5^\circ$.

\begin{figure}[htbp]
\begin{center}
\includegraphics[width=.6\textwidth]{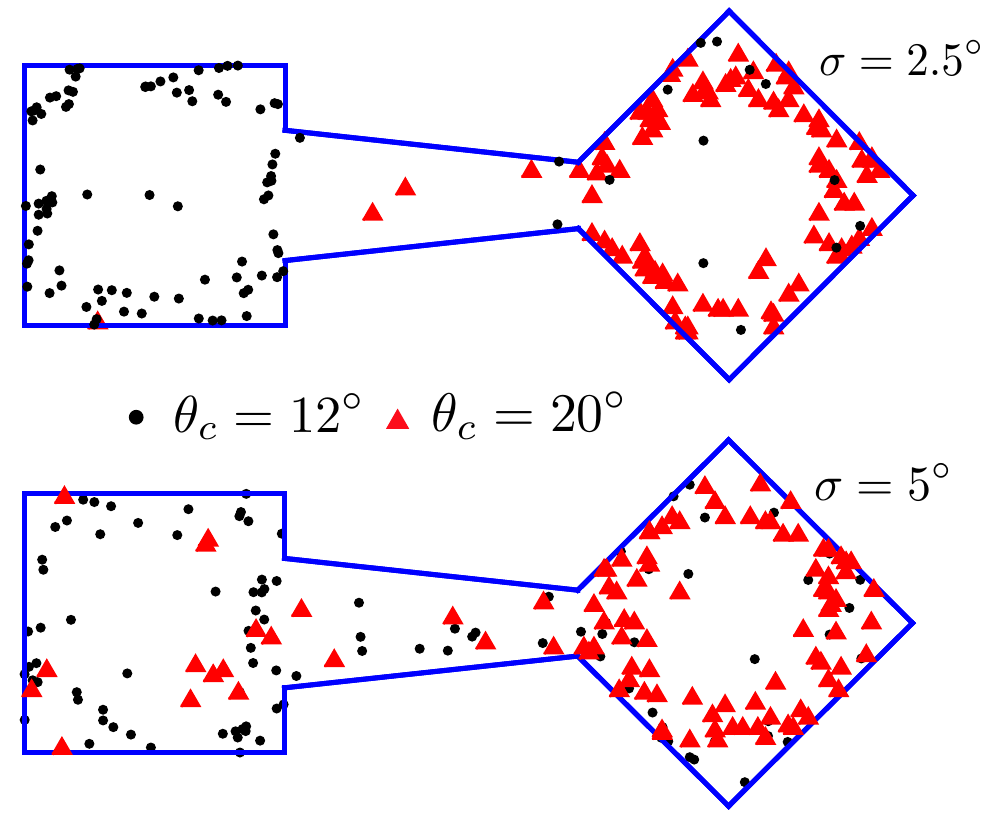}
\caption{The distribution of swimmers at $t=10$ with Gaussian perturbations of the departure angle, with standard deviation (a) $\sigma= 2.5^\circ$ and (b) $\sigma= 5^\circ$.}
\label{NoisySorting}
\end{center}
\end{figure}

\begin{figure}[htbp]
\begin{center}
\includegraphics[width=.65\textwidth]{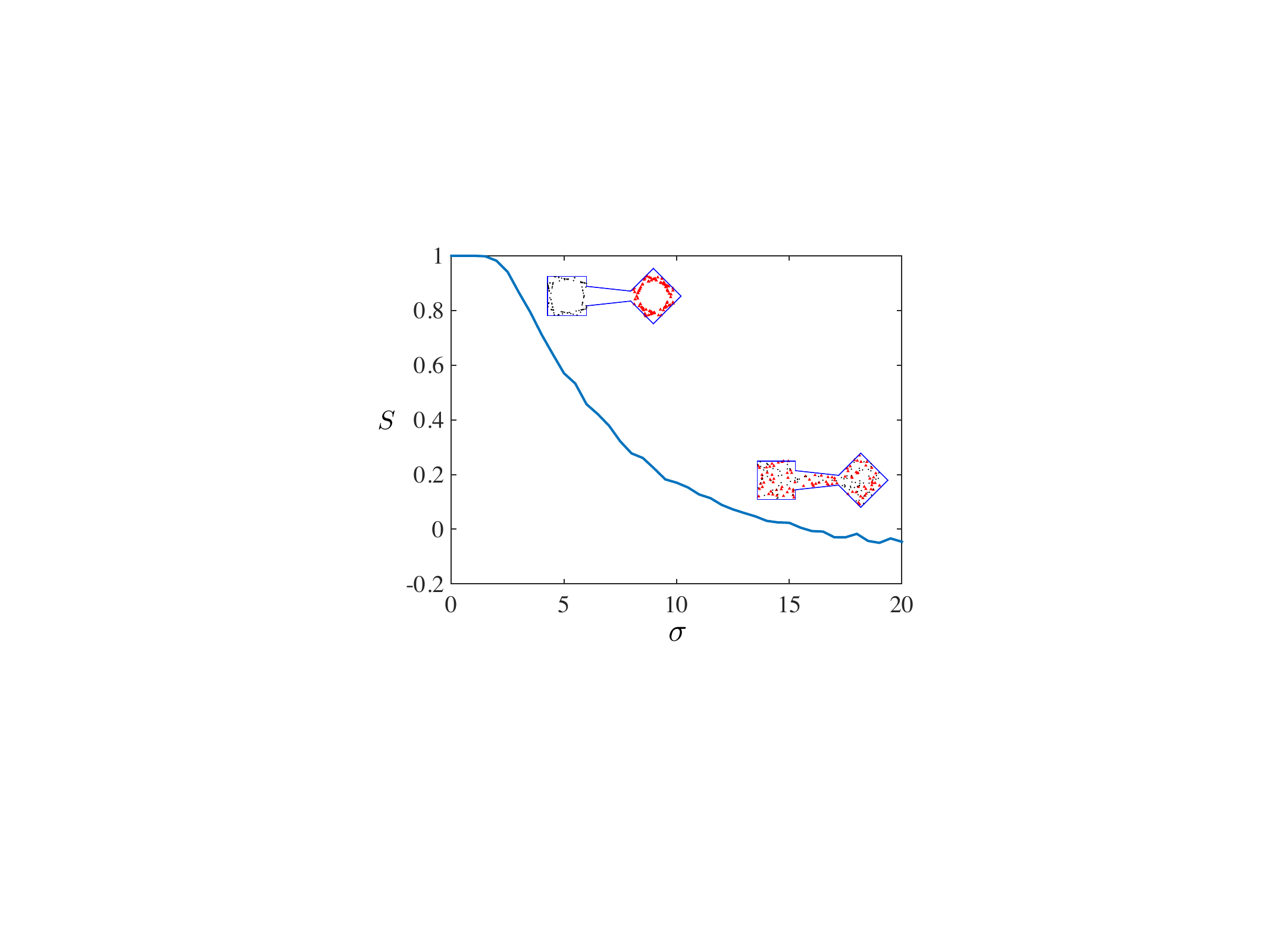}
\caption{The sorting order parameter as a function of the departure angle standard deviation. $S=1$ indicates a perfectly sorted system, while $S=0$ indicates a randomized swimmer distribution.}
\label{Order_Parameter}
\end{center}
\end{figure}

Figure~\ref{NoisySorting} shows the same sorting device but in the case that the departure angles vary with a Gaussian spread about the respective means with standard deviation $\sigma$. The swimmer distributions are shown again at $t=10$ in the cases with $\sigma=2.5^\circ$, where the sorting is significant but imperfect, and with $\sigma=5^\circ$, where the sorting is showing signs of deterioration. To measure the effectiveness of the sorting device as a function of the fluctuations in the departure angle, we consider swimmers with uniformly distributed initial positions and orientations inside the domain, and seek the probability that a swimmer is contained in the appropriate chamber after a long time. Specifically, we define an order parameter $S=(P_1+P_2)-1$, where $P_1$ is the probability that a cell with $\theta_c=12^\circ$ is contained in the leftmost chamber as $t\to \infty$ and $P_2$ is the probability that a cell with $\theta_c=20^\circ$ is contained in the rightmost chamber as $t\to \infty$. When $S=1$ the system outside of a measure zero set is eventually sorted perfectly, while $S=0$ is indicative of a disordered system. We estimate the order parameter by simulating $10,000$ randomly placed swimmers of each type and counting the number of properly sorted swimmers at $t=100$. The result is shown in Fig.~\ref{Order_Parameter}. Intuitively we have perfect sorting for small Gaussian fluctuations of the departure angles, with a steadily diminishing sorting for large fluctuations.

\section{Microorganism billiards in a periodic array of square obstacles}\label{sec: external}
\begin{figure*}[htbp]
\begin{center}
\includegraphics[width=.98\textwidth]{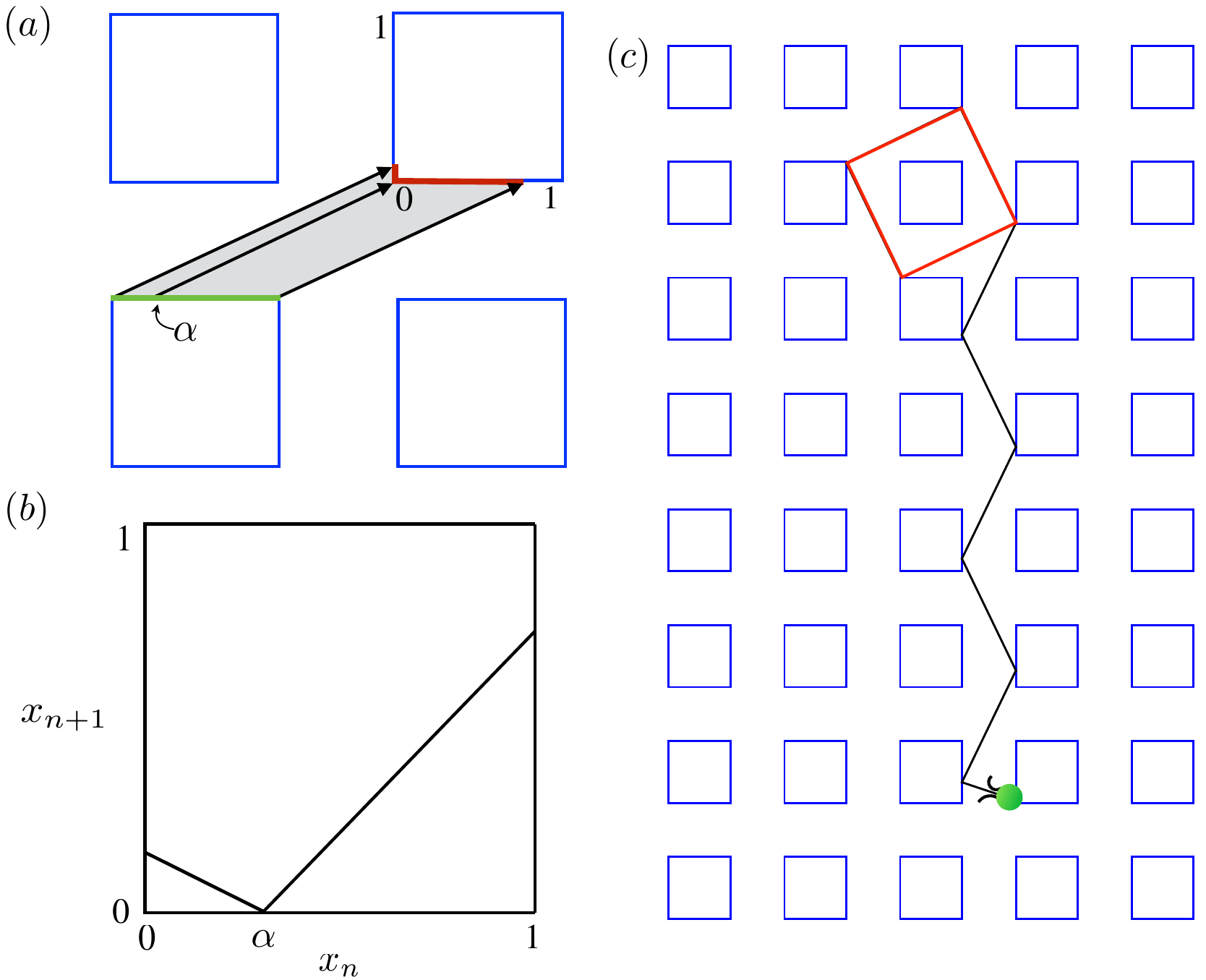}
\caption{The external microorganism billiard problem in an array of unit square obstacles. (a) Illustration of the case with obstacle distance $L=1.65$  and departure angle $\theta_c=25^\circ$. The trajectory only reaches the two visible sides of one other obstacle when leaving from any given surface. (b) The one-dimensional map of the dynamics is continuous in this case, with a stable focusing region for $x\in[0,\alpha]$ and a neutral region for $x\in[\alpha,1]$ for $\alpha=0.26$. (c) An illustrative trajectory. The swimmer is eventually trapped into a periodic orbit identical to that in the internal billiard problem in a square.}
\label{lattice_simple_map_beam_traj}
\end{center}
\end{figure*}

The approach used to study microorganism billiards inside a regular polygon can be fruitfully applied to the external problem of locomotion in a periodic array of polygonal boundaries. Recent related experiments by Volpe {\it et al.~}\cite{vbvkb11} and Brown {\it et al.~}\cite{bvdvslp15} have shown ballistic, diffusive, and entrapped dynamics of synthetic swimming Janus particles in a lattice of obstacles. Battista {\it et al.~}\cite{bfs14} studied the locomotion of malaria parasites, {\it Plasmodium sporozoites}, in an array of round pillars in a hexagonal lattice and connected stable migration to the organism's mechanical flexibility. Here we consider a swimmer in an infinite lattice of square pegs, each of unit area with centers separated by a distance $L=1.65$. The identification of each surface with every other surface once again results in a one-dimensional map that gives considerable insight into the swimming trajectories associated with a given departure angle $\theta_c$. As in the internal problem, $x_n$ denotes the position of the swimmer's $n$th reflection on any surface, where $x$ is increasing in the direction of swimming. We consider only the case where the body does not slide along the surface before departing, setting $\delta=0$, and explore three cases that are modestly representative of the myriad possibilities in the external problem.

\begin{figure*}[htbp]
\begin{center}
\includegraphics[width=.98\textwidth]{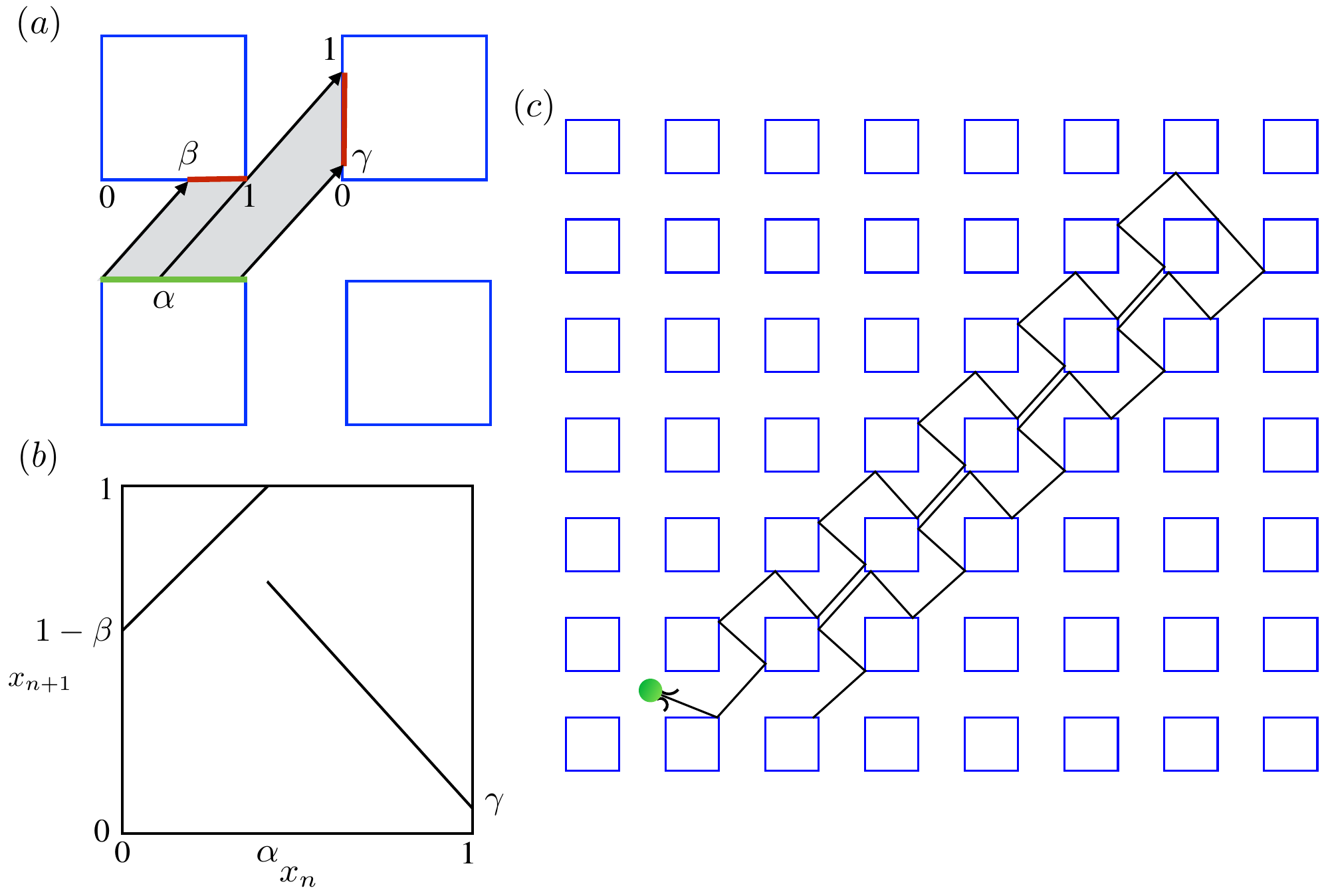}
\caption{(a) Illustration of the case $L=1.65$ and $\theta_c=48^\circ$. (b) The map contains a neutral region for $x\in[0,\alpha]$ and an unstable stretching region for $x\in[\alpha,1]$, with $\alpha=0.41$. Reflections from parallel surfaces are neutral, so unstable stretching regions also correspond to directional changes. (c) An illustrative trajectory. The dynamics jump back and forth between the neutral and unstable regions of the map, resulting in a chaotic trajectory that nevertheless retains a coherent structure. }
\label{lattice_lucy_map_beam_traj}
\end{center}
\end{figure*}

Figure~\ref{lattice_simple_map_beam_traj}a shows the first representative example, a case where the departure angle $\theta_c$ is such that a beam of swimmers departing from a given surface arrives on two surfaces of one other square obstacle. The one-dimensional map, which is continuous in this case, and a sample trajectory for $\theta_c=25^\circ$ are shown in Figs.~\ref{lattice_simple_map_beam_traj}b\&c. The map is composed of a focusing region ($|f'(x)|<1$ for $x\in[0,\alpha]$) and a neutral region ($|f'(x)|=1$ for $x>\alpha$), where $\alpha=0.26$. The result is a trajectory that is at first ballistic while the body reflects from neutral surface to neutral surface, but eventually steps into the stable focusing region, corresponding here to a periodic trajectory. The resulting orbit is identical to the orbit in the internal problem in a square domain. Generally, if $\theta_c < 45^\circ$, then the map can only contain neutral and stable regions, so that the microorganism billiard is eventually trapped into a periodic trajectory. In addition, reflections from parallel surfaces are neutral since they do not vary the length of the beam of swimmers arriving there, so any focusing or stretching dynamics must also be associated with directional changes (hitting walls that are perpendicular to the wall of departure).

A second example is shown in Figs.~\ref{lattice_lucy_map_beam_traj}a-c, where $\theta_c=48^\circ$. The swimmer can arrive at two surfaces, but now on two different obstacles. There remains a neutral region ($|f'(x)|=1$ for $x\in[0,\alpha]$) with $\alpha=0.41$. However, since $\theta_c>45^\circ$, the region $x\in[\alpha,1]$ is now stretched ($|f'(x)|>1$). The resulting trajectory therefore never settles to a stable periodic orbit. Figure~\ref{lattice_lucy_map_beam_traj}c shows a sample trajectory for this case. Although there can be no stable periodic orbit, we do observe an order to the chaotic trajectory. It is possible, for instance, for the trajectory to undergo a periodic oscillation between the neutral and unstable regions of the map, leading to a weakly unstable periodic trajectory.

\begin{figure*}[htbp]
\begin{center}
\includegraphics[width=.94\textwidth]{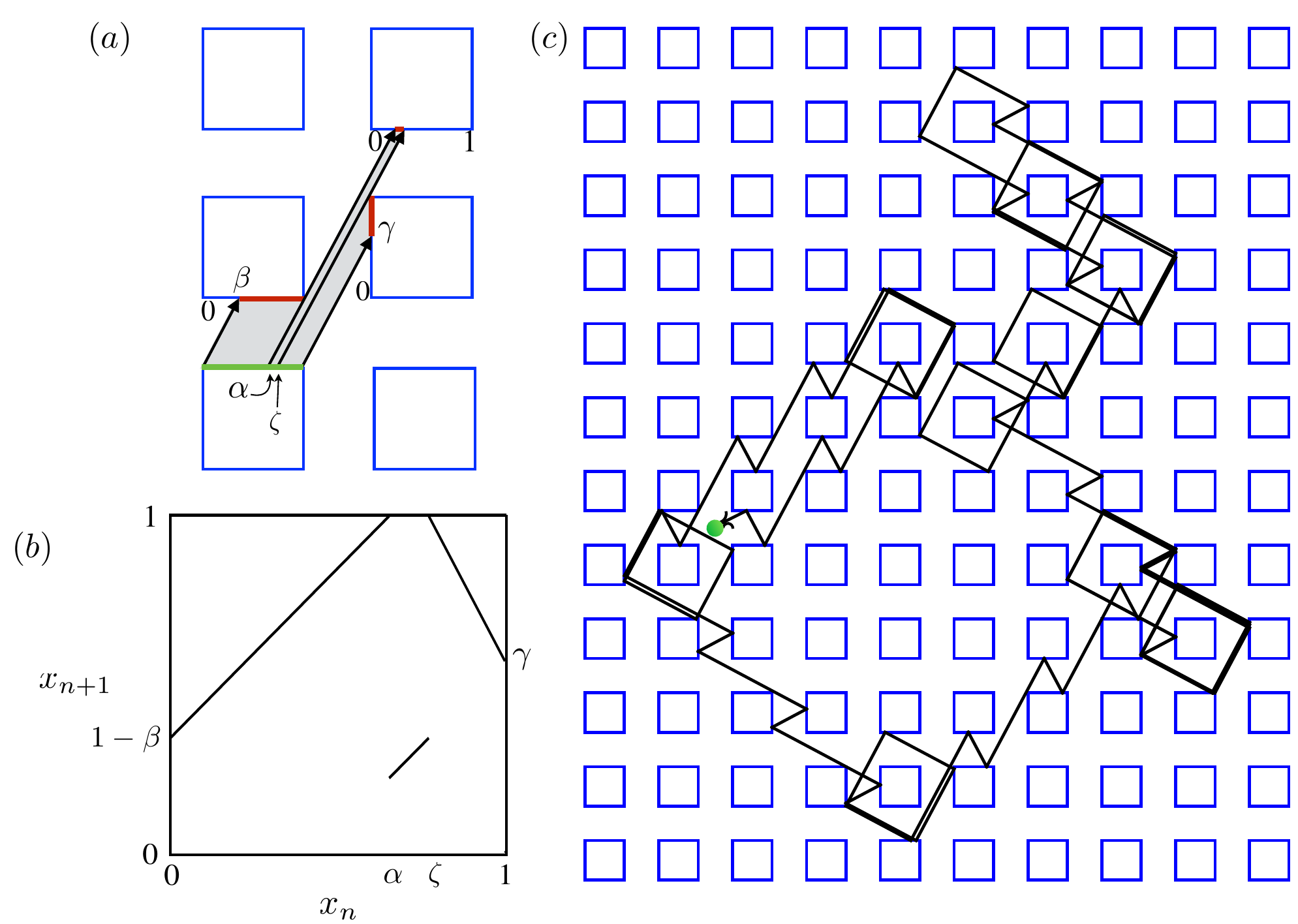}
\caption{(a) Illustration of the case $L=1.65$ and $\theta_c=62^\circ$. (b) The map contains two neutral regions corresponding to parallel wall reflections, and one unstable stretching region corresponding to directional changes. (c) An illustrative trajectory. The trajectory is chaotic; the swimmer is occasionally drawn into temporarily trapped dynamics, as in the Fig.~\ref{lattice_simple_map_beam_traj}c, but due to the swimmer sampling the {\it unstable} stretching region.}
\label{sheila_beam_traj_diff}
\end{center}
\end{figure*}

As a third and final example we set $\theta_c=62^\circ$, shown in Figs.~\ref{sheila_beam_traj_diff}a-c., with $\alpha=0.65$.  In this case the swimmer can reach three surfaces; the two surfaces parallel to the initial wall are neutral, while the surface perpendicular to the initial wall is an unstable stretching region. As in the previous example, the trajectory therefore samples both the neutral and the stretching regions, leading to a chaotic trajectory. A quasi-ballistic drift along neutral (parallel) surfaces is punctuated by occasional directional changes (perpendicular surfaces). Counter-intuitively, the swimmer is thus occasionally drawn into what appear to be almost trapped dynamics, as in the first example, but this is due to the swimmer sampling the {\it unstable} stretching region.

\section{Discussion}\label{sec: discussion}

In this paper we investigated the two-dimensional billiard-like motion of microorganisms which upon reflection from a surface depart with an angle independent of the angle of incidence. We first considered the swimming dynamics inside a regular N-sided polygon. For departure angles $\theta_c \in  (0,\pi/N)$, resulting in only consecutive wall impacts, the swimmer was found to settle into a stable periodic trajectory (an inscribed regular polygon of the same degree). This stable orbit was shown to be robust to small random fluctuations of the arrival position or the departure angle, and independent of the sliding distance $\delta$ so long as the body does not slide past a corner. What dynamics will ensue if the body does in fact slide past a corner remains unclear. Depending on the swimmer geometry and propulsive mechanism the swimming trajectory may be a simple extension of those already discussed, there may be trapping at the corner below a critical polygon interior angle, there may be continuous sliding along the entire domain \cite{skupts16}, or possibly something more exotic will appear. Though not discussed here, trapping at corners with acute opening angles is possible even without sliding \cite{cm06}. Even in classical billiards the effects of corners is an active topic of research \cite{tr03,rt12}.

For non-consecutive impacts with $\theta_c \in(\pi/N, \pi/2)$, a one-dimensional piecewise linear map may be used to characterize the reflection regions as stable (focusing), neutral, or unstable (stretching), which is an easy way to predict whether the trajectory will settle to a stable orbit or undergo unstable chaotic dynamics. The Lyapunov exponent for these maps proved to be a useful measure of chaos since there is a direct relationship between the chaos in the map and the chaos in the full system. The special case in which the angle of departure is an integer multiple of the interior angle of the polygon ($\theta_c = k\pi/N$) resulted in neutrally stable periodic orbits for any initial position. The stable fixed points derived for microorganism billiards inside a regular polygon were then used to design a model sorting device to passively separate two different types of swimmers. The effectiveness of the sorting device was investigated as a function of Gaussian variance in the departure angle. The present work may shed light on the nature of repeated wall reflections in confined and patterned domains, and may suggest novel methods for directing the transport of microorganisms. 

The one-dimensional mapping approach was also applied to the external problem of a microorganism swimming in an infinite array of square obstacles, for which we presented a few representative examples. Depending on the departure angle and the spacing between obstacles, the trajectory may become bound in a trapped orbit, or may undergo more complex and possibly chaotic dynamics. A more complete characterization of trajectories and diffusive behavior in the external problem remains an interesting future direction for study. 

To understand the behavior of larger microorganisms in confined domains will likely require more detailed modeling of the body mechanics and hydrodynamic interaction with the surface. For instance, experimental and numerical studies of swimming spermatozoa past an edge have shown that the scattering angle depends on a complex elastohydrodynamic interaction \cite{kdpg13,mgs14}. Another example is the flexible malaria parasite, {\it Plasmodium sporozoite}, which is drawn with preference to the obstacles that provide a better fit to its shape in a hexagonal array of round pillars \cite{bfs14}. Yet larger organisms such as the nematode {\it C.~elegans}, swimming in a regular array of obstacles, also exhibit complex trajectories and gait changes depending on the obstacle spacing \cite{mkzs12}. While such organisms may be hydrodynamically attracted to surfaces and drawn inexorably towards one of infinite expanse \cite{lp09,sl12}, they may slide off of the obstacles with sufficiently large gap spacing in the external problem, and therefore might still be well understood using the mapping approach presented herein. Looking forward, applications of microorganism billiard dynamics may therefore include improved spermatozoan selection to increase the success rates in in-vitro fertilization techniques, and the filtration of certain parasites for study or elimination by their mechanical properties.

  The authors thank David Anderson, Dwight Barkley, Renske Gelderloos, Michael Graham, Thomas
  Kurtz, and Douglas Weibel for helpful discussions.  This research was
  supported by NSF grants DMS-1109315 and DMS 1147523.

\appendix

\section{Exact expressions for the microorganism billiard inside a regular polygon}
\label{apx}

For a regular polygon of degree $N$, treating the plane in complex variables and the side of departure emanating from the origin in the direction $e_0=1$, the sides are parallel to the unit vectors
\begin{gather}
v_m = e^{2\pi i m/N}.
\end{gather}
Assume that $\theta_c\in(k\pi /N,(k+1)\pi/N)$ for some integer $k$. Then, writing the swimming direction as $d_c=\exp(i \theta_c)$, we have
\begin{gather}
\sum_{m=0}^k v_m +\beta v_{k+1}=\lambda_1 d_c,\ \ \sum_{m=0}^k v_m =\alpha+\lambda_2 d_c,
\end{gather}
for as yet unknown transit distances $\lambda_1,\lambda_2$. Multiplying each side in both equations by $\bar{d_c}$ and taking the imaginary parts, we find
\begin{gather}
\Im\left[e^{-i\theta_c}\left(\sum_{m=0}^k e^{2\pi im/N}+\beta e^{2\pi i(k+1)/N}\right)\right]=0,\\
\Im\left[e^{-i\theta_c}\sum_{m=0}^k e^{2\pi i m/N}\right] =-\alpha \sin(\theta_c),
\end{gather}
which, using
\begin{gather}
\sum_{m=0}^k e^{2\pi im/N}=\frac{1-\exp\left(2\pi i (k+1)/N\right)}{1-\exp\left(2\pi i/N\right)},
\end{gather}
eventually simplifies to give
\begin{gather}
\alpha=\kappa\frac{\sin \left(\theta_c-  k\pi/N\right)}{\sin (\theta_c)},\\
\beta=\kappa\frac{ \sin \left(\theta_c-k\pi  /N\right)}{ \sin \left( 2(k+1)\pi /N-\theta_c\right)}.
\end{gather}
where
\begin{gather}
\kappa=\frac{\sin \left((k+1)\pi/N\right)}{ \sin \left(\pi/N\right)}.
\end{gather}
Note that the important quantity $\beta \alpha^{-1}$ simplifies to
\begin{gather}
\beta \alpha^{-1}=\frac{\sin (\theta_c)}{\sin \left(2(k+1)\pi/N-\theta_c\right)}.
\end{gather}

\bibliographystyle{unsrt}
\bibliography{master_reference}

\end{document}